\documentclass[10pt, twoside, openright]{report}

\usepackage{preamble}

\begin{document}

\pagenumbering{gobble}
\begin{center}
\begin{figure}[h]
    \hspace*{-1cm}
    \includegraphics[width=8cm]{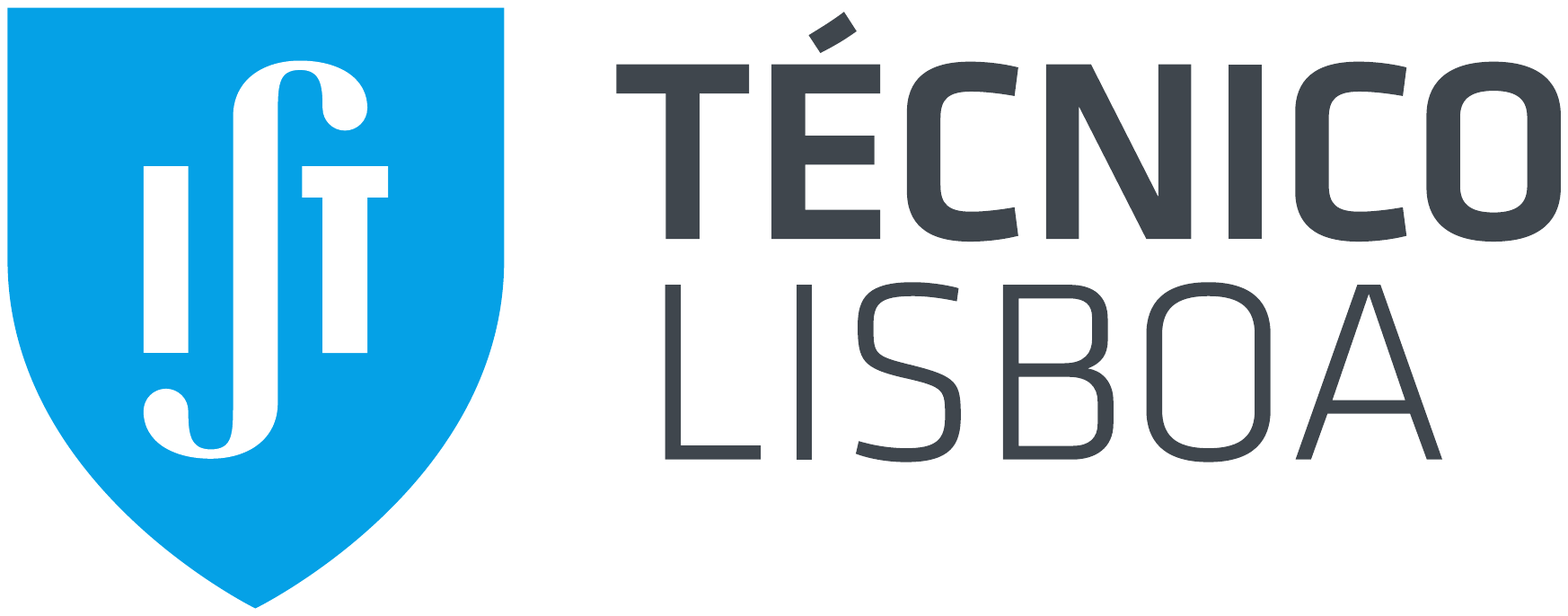}
\end{figure}

\fontsize{16pt}{0}\textbf{Library of efficient algorithms for phylogenetic analysis}
\vspace{0.5cm}

\fontsize{14pt}{0}\textbf{Luana Bernardino da Silva}
\vspace{2cm}

\fontsize{12pt}{0}\selectfont Thesis to obtain the Master of Science Degree in
\vspace{0.25cm}

\fontsize{16pt}{0}\textbf{Computer Science and Engineering}
\vspace{0.5cm}

\fontsize{12pt}{0}\selectfont Supervisors: Doctor Alexandre Paulo Lourenço Francisco
\vspace{0.25cm}

\fontsize{12pt}{0}\selectfont Doctor Cátia Raquel Jesus Vaz
\vspace{2cm}

\fontsize{14pt}{0}\textbf{Examination Committee}
\vspace{0.5cm}

\fontsize{12pt}{0}\selectfont Provisory
\vspace{0.25cm}

\vspace{5.25cm}
\fontsize{14pt}{0}\textbf{December 2020}
\end{center}
\cleardoublepage
\pagenumbering{roman}
\chapter*{Acknowledgements}

This thesis has been both a pleasure and a challenge to accomplish, and it would not have been made possible without the tremendous help and support I have received.

I would first like to thank my supervisors, Alexandre Francisco and Cátia Vaz, for providing me with the valuable support and guidance needed to accomplish this thesis, and most importantly for even giving me the opportunity to be a part of it.

I would also like to thank my partner for all the help and motivation he gave me to keep going despite all of the challenges, and for always being there for me through all the ups and downs.

Finally, I would like to thank my family and friends for believing in me, and for always providing me fun distractions that create happy memories to help me keep going when times get rough.




\vfill
\noindent This work was partly supported by national funds through FCT -- Fundação para
a Ciência e Tecnologia, under projects PTDC/CCI-BIO/29676/2017 and
UIDB/50021/2020.
\chapter*{Abstract}

Evolutionary relationships between species are usually inferred through phylogenetic analysis, which provides phylogenetic trees computed from allelic profiles built by sequencing specific regions of the sequences and abstracting them to categorical indexes. With growing exchanges of people and merchandise, epidemics have become increasingly important, and combining information of country-specific datasets can now reveal unknown spreading patterns.

The phylogenetic analysis workflow is mainly composed of four consecutive steps, the distance calculation, distance correction, inference algorithm, and local optimization steps. There are many phylogenetic tools out there, however most implement only some of these steps and serve only one single purpose, such as one type of algorithm. Another problem with these is that they are often hard to use and integrate, since each of them has its own API.

This project resulted a library that implements the four steps of the workflow, and is highly performant, extensible, reusable, and portable, while providing common APIs and documentation. It also differs from other tools in the sense that, it is able to stop and resume the workflow whenever the user desires, and it was built to be continuously extended and not just serve a single purpose.

The time benchmarks conducted on this library show that its implementations of the algorithms conform to their theoretical time complexity. Meanwhile, the memory benchmarks showcase that the implementations of the NJ algorithms follow a linear memory complexity, while the implementations of the MST and GCP algorithms follow a quadratic memory complexity.

\vspace{0.5cm}

\textbf{Key words:} phylogeny; sequences; profiles; inference; algorithms; trees.

\chapter*{Resumo}

As relações evolucionárias entre diferentes espécies são geralmente inferidas através de análise filogenética, que fornece árvores filogenéticas, que podem ser computadas através de perfis alélicos construídos sequenciando regiões específicas das sequências e abstraíndo-as em índices categóricos. Com o aumento de trocas de pessoas e mercadorias, as epidemias têm-se tornado muito importantes, e combinar informações de datasets específicos por país pode agora revelar padrões de propagação desconhecidos.

O fluxo de análise filogenética é composto principalmente por quatro passos consecutivos, o cálculo de distâncias, a correção de distâncias, o algoritmo de inferência, e a otimização local. Existem muitas ferramentas de filogenia, porém muitas implementam apenas alguns destes passos e servem apenas um propósito, por exemplo um tipo de algoritmos. Outro problema é que muitas vezes são difíceis de usar e integrar, porque cada uma tem a sua API.  

Este projeto resultou numa biblioteca que implementa os quatro passos do fluxo, é eficiente, extensível, reutilizável, e portável, e fornece APIs comuns e documentação. Esta difere das outras no sentido em que, é capaz de parar e resumir o fluxo sempre que o utilizador deseja, e foi construída para ser continuamente estendida e não servir apenas um propósito.

Os benchmarks de tempo conduzidos sobre esta biblioteca mostram que as suas implementações dos algoritmos estão conforme as suas complexidades de tempo teóricas. Os benchmarks de memória demonstram que as implementações dos algoritmos de NJ seguem uma complexidade de memória linear, enquanto que as implementações dos algoritmos de MST e GCP seguem uma complexidade de memória quadrática.

\vspace{0.5cm}

\textbf{Palavras Chave:} filogenia; sequências; perfis; inferência; algoritmos; árvores.

\tableofcontents
\addcontentsline{toc}{chapter}{\listfigurename}
\listoffigures
\addcontentsline{toc}{chapter}{\listtablename}
\listoftables
\addcontentsline{toc}{chapter}{List of Acronyms}
\chapter*{List of Acronyms}

\begin{acronym}[MPC]
    \acro{DNA}{deoxyribonucleic acid}
    \acro{HPC}{High Performance Computing}
    \acro{NGS}{Next Generation Sequencing}
    \acro{EBNF}{Extended Backus–Naur Form}
    \acro{MLST}{Multilocus Sequence Typing}
    \acro{MLVA}{Multiple-Locus Variable Number Tandem Repeat Analysis}
    \acro{CSV}{comma-separated values}
    \acro{SNP}{Single Nucleotide Polymorphism}
    \acro{GCP}{Globally Closest Pairs}
    \acro{ME}{Minimum Evolution}
    \acro{NJ}{Neighbour Joining}
    \acro{MST}{Minimum Spanning Tree}
    \acro{SL}{Single-linkage}
    \acro{CL}{Complete-linkage}
    \acro{UPGMA}{Unweighted Pair Group Method with Arithmetic-mean}
    \acro{UPGMC}{Unweighted Pair Group Method with Centroid}
    \acro{WPGMA}{Weighted Pair Group Method with Arithmetic-mean}
    \acro{WPGMC}{Weighted Pair Group Method with Centroid}
    \acro{UNJ}{Unweighted Neighbour Joining}
    \acro{FNJ}{Fast Neighbour Joining}
    \acro{RNJ}{Relaxed Neighbour Joining}
    \acro{goeBURST}{globally optimized eBURST}
    \acro{eBURST}{eletronic Based Upon Related Sequence Types}
    \acro{ST}{Sequence Type}
    \acro{SLV}{Single Locus Variant}
    \acro{DLV}{Double Locus Variant}
    \acro{TLV}{Triple Locus Variant}
    \acro{cgMLST}{core genome MLST}
    \acro{LBR}{Local Branch Recrafting}
    \acro{SPR}{Subtree Pruning and Regrafting}
    \acro{NNI}{Nearest Neighbor Interchange}
    \acro{TBR}{Tree Bisection and Reconnection}
    \acro{JVM}{Java Virtual Machine}
    \acro{UML}{Unified Modeling Language}
    \acro{CLI}{Command Line Interface}
\end{acronym}

\cleardoublepage
\pagenumbering{arabic}
\chapter{Introduction}

The evolutionary relationships between different species or taxa are usually inferred through known phylogenetic analysis techniques. Some of these techniques rely on the inference of phylogenetic trees, which can be computed from \ac{DNA} sequences, or from allelic profiles built by sequencing specific regions of the sequences and abstracting them to categorical indexes. Phylogenetic trees are also used in other contexts, such as to understand the evolutionary history of gene families, to allow phylogenetic foot-printing, to trace the origin and transmission of infectious diseases, or to study the co-evolution of hosts and parasites \cite{inference}.

With growing exchanges of people and merchandise between countries, epidemics have become an issue of increasing importance, thus epidemiological surveillance is now a global procedure rather than a country-based one. Combining information of country-specific datasets can now reveal epidemic spreading patterns that were not possible to detect before, but phylogenetic algorithms are often hard to use and integrate in analysis frameworks and tools.

There are hundreds of computational phylogenetics tools out there that are commonly used to address this problem. Although they all try to achieve the same goal, which is to build a phylogenetic tree, they all differ widely in the way they operate, the formats they support, and the criteria and algorithms they implement. Due to those differences, the use of different tools may result in different phylogenetic trees \cite{trees} from the same algorithm. There is not yet a library that tries to integrate all of the algorithms into just one library, and that works on all platforms and can be integrated with other tools.

\begin{figure}[!ht]
    \centering
    \includegraphics[scale=0.7]{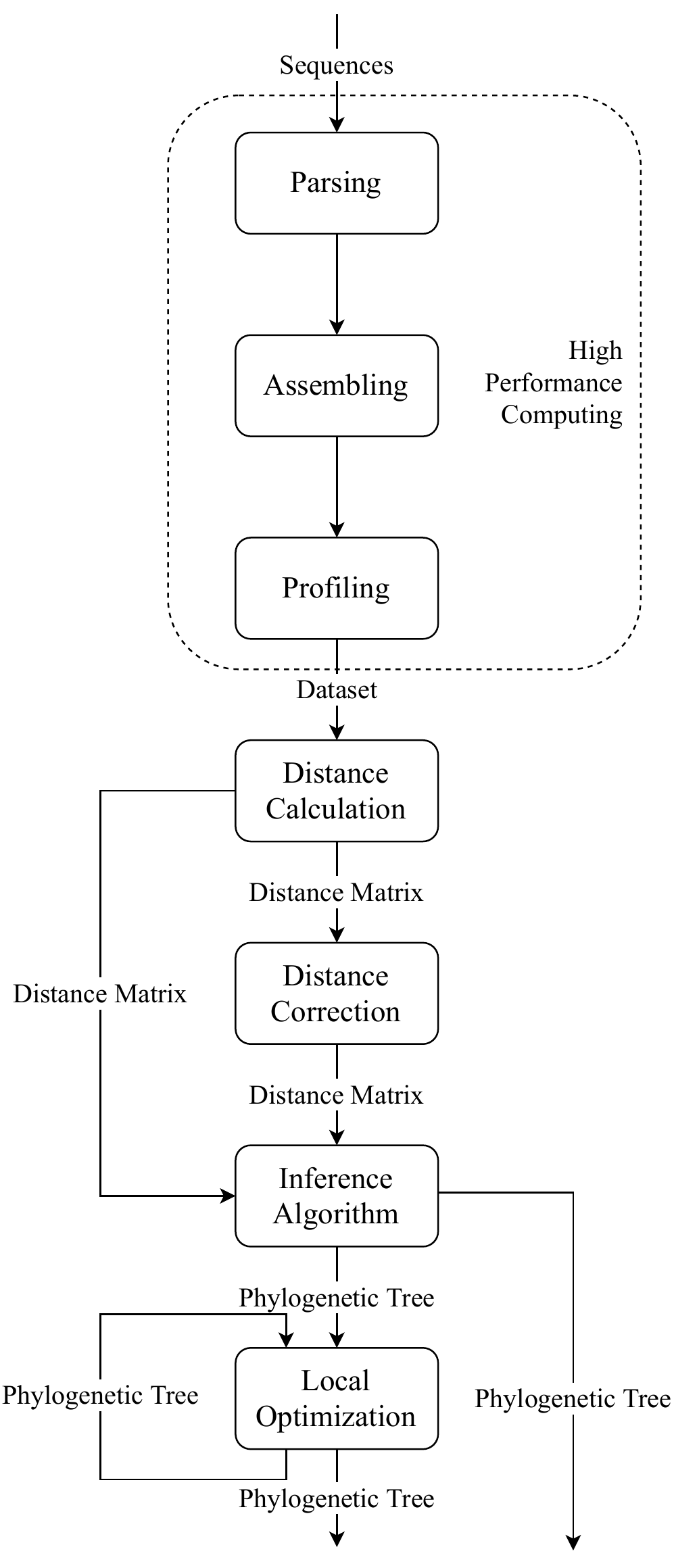}
    \caption{Workflow abstraction of the phylogenetic analysis in the INNUENDO project.}
    \label{fig:placement}
\end{figure}

The process of phylogenetic analysis consists of parsing, assembling, and profiling the sequences, so that they can then be processed by a distance calculation metric and an optional distance correction metric, followed by an inference algorithm and multiple optional local optimizations \cite{distance}. This process is exposed in projects like INNUENDO \cite{innuendo}, which performs these operations in High Performance Computing (HPC) \cite{hpc} pipelines. However, most of them, like INNUENDO, do not compute the steps after the parsing, assembling and profiling of sequences in the pipelines, because there is not yet a library that can be integrated to compute those parts. Figure \ref{fig:placement} demonstrates an abstraction of the workflow of the INNUENDO project, regarding the \ac{HPC} pipeline placement.

\section{Objectives}

The aim of this project is the development of a library of phylogenetic algorithms and related data structures, with suitable common APIs and readily available documentation. The suitable common APIs should allow the library to be easily extended to include other algorithms and formats. And they should also allow the user to better explore the differences between algorithms through the outputs. This project will rely on already existing algorithm prototypes and on ongoing research work at INESC-ID and iMM. The resulting library will be tested and integrated in the INNUENDO project for large scale \ac{NGS} \cite{ngs} data analysis and in tools such as PHYLOViZ \cite{phyloviz}.

\section{Document Structure}

This document starts by explaining, in the Background chapter, the concepts related to this project, in terms of phylogeny, similarity, clustering, and optimization, and by then mentioning and comparing some of the already existing related work. The Proposed Solution chapter defines the functional and non functional requirements, as well as use cases for this project. It also contains the architecture definition and the choice of technologies to be used. The Implementation chapter describes in more depth how the proposed solution was implemented, focusing on the most important aspects of the implementation. The Experimental Evaluation chapter analyzes the results obtained for the running time and memory usage of each algorithm implementation. Lastly, the Final Remarks chapter summarizes the most important points of this thesis and enumerates possible future work to extend and improve on this project. This project is publicly available at \url{https://github.com/Luanab/phylolib} along with its documentation.

\chapter{Background}

This chapter is mostly grounded on the fundamentals found in the article \textit{Large Scale and Dynamic Phylogenetic Inference from Epidemic Data} \cite{martanascimento} by Marta Nascimento. For a more in-depth reading on the topics that will be discussed here please refer to that article.

In biology, phylogenetics is the study of the evolutionary history and relationships among individuals or groups of organisms (e.g. species, or populations). These relationships are discovered through phylogenetic inference algorithms that evaluate observed heritable traits, such as \ac{DNA} sequences or morphology under a model of evolution. These models try to describe the evolution process of the species from which a sequence of symbols changes into another set of traits, and differ in terms of the parameters used to describe the rates at which one nucleotide replaces another during evolution. For instance, they are used during the calculation of likelihood of a tree or to estimate the evolutionary distance between sequences from the observed differences. This enables us to infer evolutionary events that happened in the past, and also provides more information about the evolutionary processes operating on sequences.

\section{Phylogenetic Analysis}

Phylogenetic analysis aims at uncovering the evolutionary relationships between different species, or even between individuals of the same species, to obtain an understanding of their evolution. The result of this analysis is a phylogeny, which can be a phylogenetic tree or network, that is a diagrammatic hypothesis about the history of the evolutionary relationships of a group of organisms. The tips of a phylogeny can be living organisms or fossils, and represent the ``end'', or the present, in an evolutionary lineage.

\subsection{Analysis}

Phylogenetic analysis has become central to understanding biodiversity, evolution, ecology, and genomes. Phylogenetic trees are widely used to address this task and are reconstructed by several different algorithms. They are a subset of phylogenetic networks, where nodes can only have one parent instead of two. However, a phylogenetic tree will not always be enough to correctly represent the evolutionary history of a population and sometimes a network representation will be more appropriate. Phylogenetic networks provide an alternative to phylogenetic trees and may be more suitable for datasets whose evolution involves significant amounts of reticulate events caused by hybridization, horizontal gene transfer, recombination, gene conversion or gene duplication and loss. However, they are hard to analyze and thus phylogenetic trees are more used. Therefore, this project will focus itself on phylogenetic trees and commonly used algorithms to reconstruct them.

A phylogenetic tree can be rooted or unrooted. A rooted tree is a dendrogram that indicates the common ancestor, or ancestral lineage, of the tree. An example of this type of tree is present in Figure \ref{fig:rooted}. An unrooted tree however makes no assumption about the ancestral line, and does not show the origin or ``root" of the gene or organism in question. An example of this type of tree is present in Figure~\ref{fig:unrooted}.

\begin{figure}[!ht]
    \centering
    \begin{minipage}{.4\textwidth}
        \centering
        \begin{tikzpicture}
			\node (a) at (0.5, 0) {A};
			\node (b) at (1.5, 0) {B};
			\node (c) at (2.5, 0) {C};
			\node (d) at (3.5, 0) {D};
			\node (e) at (4.5, 0) {E};
			\node (ab) at (1, 1) {};
			\node (cd) at (3, 1.5) {};
			\node (abcd) at (2, 2.5) {};
			\node (all) at (2, 3.5) {};
			
    		\node (1) at (0.1, 0.6) {1.3};	
    		\node (2) at (2.1, 0.8) {1.9};	
    		\node (3) at (0.7, 1.7) {2};	
    		\node (4) at (4.9, 1.8) {4.1};
    		
			\draw  (a) |- (ab.center);
			\draw  (b) |- (ab.center);
			\draw  (c) |- (cd.center);
			\draw  (d) |- (cd.center);
			\draw  (cd.center) |- (abcd.center);
			\draw  (ab.center) |- (abcd.center);
			\draw  (abcd.center) |- (all);
			\draw  (e) |- (all);
		\end{tikzpicture}
        \captionof{figure}{Example of a rooted phylogenetic tree.}
        \label{fig:rooted}
    \end{minipage}
    \hspace{.1\textwidth}
    \begin{minipage}{.4\textwidth}
        \centering
        \begin{tikzpicture}
    		\node (a) at (0, 3) {A};
    		\node (b) at (1, 4.7) {B};
    		\node (c) at (4, 5) {C};
    		\node (d) at (3.8, 3) {D};
    		\node (e) at (2.5, 1.7) {E};
    		\node (ab) at (1.25, 3.8) {};
    		\node (cd) at (3.5, 3.8) {};
    		\node (abcd) at (2.5, 3.5) {};
        	
    		\node (1) at (0.6, 3.7) {10};	
    		\node (2) at (1.5, 4.2) {5.3};	
    		\node (3) at (1.8, 3.4) {9};	
    		\node (4) at (2.9, 3.9) {7};	
    		\node (5) at (4, 4.2) {11};	
    		\node (6) at (3.9, 3.6) {5};	
    		\node (7) at (2.9, 2.7) {12.5};
    		
    		\draw (a)-\\(ab.center);
    		\draw (b)-\\(ab.center);
    		\draw (c)-\\(cd.center);
    		\draw (d)-\\(cd.center);
    		\draw (ab.center)-\\(abcd.center);
    		\draw (cd.center)-\\(abcd.center);
    		\draw (e)-\\(abcd.center);
		\end{tikzpicture}
        \captionof{figure}{Example of an unrooted phylogenetic tree.}
        \label{fig:unrooted}
    \end{minipage}
\end{figure}

\subsection{Data Formats}

There are two commonly used text formats for representing phylogenetic trees, which are Newick \cite{newick} and Nexus \cite{nexus}.

Newick is a format where: each node is represented by an id and weight separated by a colon; siblings are also separated by a comma; children are enclosed in parentheses; and internal nodes are represented like any other node, except for the id that is omitted. Having in mind the following definitions:
\begin{verbatim}
    Tree: the full input Newick Format for a single tree.
    Subtree: a leaf node or an internal node and its descendants.
    Leaf: a node with no descendants.
    Internal: a node and its one or more descendants.
    BranchSet: a set of one or more Branches
    Branch: a tree edge and its descendant subtree.
    Name: the name of a node.
    Length: the length of a tree edge.
\end{verbatim}
it is possible to define the full grammar rules for Newick in \ac{EBNF} \cite{ebnf} as follows:
\begin{verbatim}
    Tree = Subtree , ";" | Branch , ";";
    Subtree = Leaf | Internal;
    Leaf = Name;
    Internal = "(" , BranchSet , ")" , Name;
    BranchSet = Branch | Branch , "," , BranchSet;
    Branch = Subtree Length;
    Name = empty | string;
    Length = empty | ":" , number;
\end{verbatim}

The Newick representation in Figure \ref{fig:newick} for the tree presented in Figure \ref{fig:rooted} can be obtained by following the previous rules.
\begin{figure}[!ht]
\centering
\begin{minipage}{.35\textwidth}
\begin{verbatim}
(((A,B):1.3,(C,D):1.9):2,E):4.1
\end{verbatim}
\end{minipage}
\caption{Example of a Newick representation of Figure \ref{fig:rooted}.}
\label{fig:newick}
\end{figure}

Nexus is a format that uses headers for more detailed information about each sequence, and the Newick format for the tree representation. Each header starts with \verb|BEGIN [name];| and ends with \verb|END;|. Figure \ref{fig:nexus} presents a Nexus representation of the same tree presented in Newick format in Figure \ref{fig:newick}.
\begin{figure}[!ht]
\centering
\begin{minipage}{.5\textwidth}
\begin{verbatim}
BEGIN TAXA;
    Dimensions NTax=5;
    TaxLabels A B C D E;
END;

BEGIN CHARACTERS;
    Dimensions NChar=20;
    Format DataType=DNA;
    Matrix
    A   ACATA GAGGG TACCT CTAAG
    B   ACATA GAGGG TACCT CTAAG
    C   ACATA GAGGG TACCT CTAAG
    D   ACATA GAGGG TACCT CTAAG
    E   ACATA GAGGG TACCT CTAAG
END;

BEGIN TREES;
    Tree best=(((A,B):1.3,(C,D):1.9):2,E):4.1;
END;
\end{verbatim}
\end{minipage}
\caption{Example of a Nexus representation of Figure \ref{fig:rooted}.}
\label{fig:nexus}
\end{figure}

\section{Similarity}

The goal of phylogenetic analysis is to discover relationships between species or populations by grouping them based on some similarity criterion that underlies some evolution model.

\subsection{Typing}

The similarity criterion is applied to the strains given by the chosen typing method. The concept of typing is designated as the identification of the genome strain. Typing methods based on sequences represent strains by character states (e.g. adenine (A), cytosine (C), guanine (G), thymine (T), or gap in the case of multiple alignment of nucleotide sequences). There are some formats for representing strains, although this project will focus itself only on \ac{MLST}, \ac{MLVA}, FASTA, and \ac{SNP}.

In the FASTA format, each sequence is represented by a line with the greater-than symbol, followed by a summary description of the sequence, and then another line with the actual sequence of character states. An example of this format can be found in Figure \ref{fig:fasta}.
\begin{figure}[!ht]
\centering
\begin{minipage}{.7\textwidth}
\begin{verbatim}
> Sequence 1
GAAGCGAGTGACTTGGCAGAAACAGTGGCCAATATTCGTCGCTACCAGATGTTTGGCATC
GCGCGCTTGATTGGTGCGGTTAATACGGTTGTCAATGAGAATGGCAATTTAATTGGATAT
> Sequence 2
GAACCGAGTGACTTGGCAGAAACAGTGGCCAATATTCGTCGCTACCAGATGTTTGGCATC
GCGCGCTTGATTGGTGCGGTTAATACGGTTGTCAATGAGAATGGCAATTTAATTGGATAT
\end{verbatim}
\end{minipage}
\caption{Example of two sequences in FASTA format.}
\label{fig:fasta}
\end{figure}

The \ac{SNP} format represents each sequence by a line with a sequence of 1's and 0's preceded by a number that identifies the sequence. A value of 0 in any location represents the character state that was mostly found on that location, while a 1 represents any other possible character state. An example of this format can be found in Figure \ref{fig:snp}.
\begin{figure}[!ht]
\centering
\begin{minipage}{.7\textwidth}
\begin{verbatim}
1 0100000111101010001000101010101001010100011101011000101010
2 1111010010101100100101010101001000001010010001010101010100
\end{verbatim}
\end{minipage}
\caption{Example of two sequences in SNP format.}
\label{fig:snp}
\end{figure}

The \ac{MLST} format is based on the \ac{CSV} format with tab separators, where the first line represents the headers, and the rest represent a sequence each. In this format, the first column is a number that identifies the sequence, and the other columns are numbers that identify the alleles present in specific loci of the DNA sequence. An example of this format can be found in Figure \ref{fig:mlst}.
\begin{figure}[!ht]
$$
\begin{matrix}
ST & cox1 & rnl \\
1 & 1 & 1 \\
2 & 2 & 2 \\
3 & 3 & 2
\end{matrix}
$$
\caption{Example of three sequences in MLST format. Columns \textit{cox1} and \textit{rnl} represent the loci of the sequences, and the corresponding numbers represent the ids of the alleles observed in those loci.}
\label{fig:mlst}
\end{figure}

The \ac{MLVA} format is very similar to the \ac{MLST} format, in the sense that both are represented by numbers and each sequence corresponds to one line, with the difference that it does not have a headers line. An example of this format can be found in Figure \ref{fig:mlva}.
\begin{figure}[!ht]
$$
\begin{matrix}
15 & 7 & 14 \\
32 & 13 & 22 \\
34 & 23 & 42
\end{matrix}
$$
\caption{Example of three sequences in MLVA format.}
\label{fig:mlva}
\end{figure}

\subsection{Criterion}

Phylogenetic trees can be built using distance matrix methods or character-state methods. Distance matrix methods infer the relationship between individuals as the number of genetic differences between pairs of sequences, whereas in character-state methods is used an array of character states. This project will focus itself on distance-based analysis of \ac{DNA} sequences.

The most commonly used similarity criterion between pairs of sequences is based on the Hamming distance \cite{hamming}, defined as the proportion of positions at which two aligned sequences A and B differ, as shown in Equation \ref{eq:hamming}.
\begin{equation}\label{eq:hamming}
    D_{ij} = \sum_{l \in L} 1_{\left\{\pi_{l}(i) \neq \pi_{l}(j)\right\}}
\end{equation}

However, this distance handles missing values as normal values. To handle missing data correctly, the GrapeTree algorithm \cite{grapetree}, explained further ahead, implements a directional measure based on normalized asymmetric Hamming distances. This approach assumes that one of the sequences is the ancestor of the other and treats missing data as deletions from the ancestor to the descendant. This measure is shown in Equation \ref{eq:grapetreedistance}, where 0 is assumed to be a missing value.
\begin{equation}\label{eq:grapetreedistance}
     D_{ij} = \frac{\sum_{l \in L} 1_{\left\{\left(\pi_{l}(i) \neq \pi_{l}(j)\right) \wedge\left(\pi_{l}(j) \neq 0\right)\right\}}}{\sum_{l \in L}1_{\left\{\pi_{l}(j) \neq 0\right\}}}
\end{equation}

The Hamming distance also does not take into consideration the number of mutations that occurred at the same position and therefore it underestimates the true evolutionary distance. To rectify this, a correction formula based on some model of evolution is often used. An example is the Jukes-Cantor model \cite{jukescantor}, that assumes all substitutions are independent, sequence positions are equally subject to change, substitutions occur randomly among the four types of nucleotides, and no insertions or deletions have occurred. This can be translated into Equation \ref{eq:jukescantor}, where $H_{ij}$ is the Hamming distance given by one of the previous equations.
\begin{equation}\label{eq:jukescantor}
    D_{ij} = -\frac{3}{4} \cdot \ln \left(1-\frac{4}{3} \cdot H_{ij}\right)
\end{equation}

In the Jukes-Cantor model the mutant is chosen with equal probability among the three possible nucleotides. Kimura later modified this equation to accommodate the fact that transition events ($A \leftrightarrow G$ and $C \leftrightarrow T$) occur at a faster rate than all other events. He provided a method for inferring evolutionary distance in which transitions and transversions are treated separately. Equation \ref{eq:kimura} defines this model, where $P$ is the fraction of sequence positions differing by a transition and $Q$ is the fraction of sequence positions differing by a transversion.
\begin{equation}\label{eq:kimura}
    D_{ij} = -\frac{1}{2} \cdot \ln \left((1 - 2 \cdot P - Q) \cdot \sqrt{1 - 2 \cdot Q}\right)
\end{equation}

Both models are unrealistic in terms of all nucleotides being expected to occur with the same frequency in a random sequence, which is not likely to be the case for any sequence. Therefore, more sophisticated models have been introduced to deal with subtle differences in substitution rates, such as the models from Felsenstein \cite{felsenstein} or Hasegawa \cite{hasegawa}.

An example of a distance matrix resultant from applying the Hamming distance in Equation \ref{eq:hamming}, to the \ac{MLST} dataset in Figure \ref{fig:mlst} is presented in Figure \ref{fig:matrix}.
\begin{figure}[!ht]
$$D = 
\begin{bmatrix}
0 & 2 & 2 \\
2 & 0 & 1 \\
2 & 1 & 0
\end{bmatrix}
$$
\caption{Example of a distance matrix resultant from applying the Hamming distance in Equation \ref{eq:hamming}, to the MLST dataset in Figure \ref{fig:mlst}.}
\label{fig:matrix}
\end{figure}

This distance matrix can be represented in an asymmetric format, also known as a square format \cite{phylip}, similar to the one in the given example, with the difference that it is preceded by a line with the number of profiles, which is equal to the number of lines and columns, and each line is preceded by the profile id. An example of this format can be seen in Figure \ref{fig:square}.
\begin{figure}[!ht]
$$
\begin{matrix}
3 \\
1 & 0 & 2 & 2 \\
2 & 2 & 0 & 1 \\
3 & 2 & 1 & 0
\end{matrix}
$$
\caption{Example of a distance matrix in an asymmetric format.}
\label{fig:square}
\end{figure}

It is also possible to further simplify the previous format for the given example, since the distance matrix is symmetric, by removing duplicate and unnecessary distances, such as duplicate distances between the same two profiles and unnecessary distances to the profiles themselves. This will result in a symmetric format, also known as a lower-triangle format \cite{phylip}, similar to the previous format, but without the zeros and following values, as seen in Figure \ref{fig:triangle}.
\begin{figure}[!ht]
$$
\begin{matrix}
3 \\
1 \\
2 & 2 \\
3 & 2 & 1
\end{matrix}
$$
\caption{Example of a distance matrix in a symmetric format.}
\label{fig:triangle}
\end{figure}
\section{Clustering}

There are several algorithms that construct phylogenetic trees and they can all be seen as clustering algorithms because they apply several clustering techniques in their approach. Clustering is an unsupervised learning problem. Given a set of elements the goal is to group them in such a way that elements in the same group (called a cluster) are more related (similar) to each other than to those in other groups (clusters). It is a main task of exploratory data mining, and a common technique for statistical data analysis, used in many fields, including machine learning, pattern recognition, image analysis, information retrieval, bioinformatics, data compression, and computer graphics. Clustering can be divided into two types: hierarchical clustering and flat (or partitioning) clustering.

Hierarchical clustering seeks to build a hierarchy of clusters. There are two techniques used to build the hierarchy: agglomerative and divisive. Agglomerative is a “bottom up” approach where each element is in its own cluster and, as the hierarchy moves up, a pair of clusters is merged into one. Divisive clustering is a “top-down” approach where all elements are together in one cluster and, as the hierarchy moves down, a cluster is split in two.

Flat clustering tries to build a group of clusters all independent from each other (i.e. there is no relation among them) and can also be divided in two categories: hard and soft clustering. Hard clustering computes a hard assignment, where each element is a member of exactly one cluster, while the latter computes a soft assignment, where each element’s assignment is a probability distribution over all clusters (i.e. an element can belong to several clusters).

This project will focus itself on hierarchical agglomerative clustering to build the hierarchy. This type of clustering has been extensively used in bioinformatics and computational biology, namely in phylogenetic inference within most phylogenetic tree reconstruction algorithms.

Distance-based hierarchical agglomerative clustering algorithms may be based on \ac{GCP}, by starting with the most similar sequences, or \ac{ME} principle, by trying to minimize the total branch length of the tree. Algorithms based on \ac{ME} can descend from \ac{NJ} or \ac{MST} algorithms.

\paragraph{Generalization}

All algorithms based on \ac{GCP}, \ac{NJ} or \ac{MST} follow a general scheme that is represented in Algorithm \ref{alg:pseudocode}, which receives as input a distance matrix containing all pairwise distances between elements and returns a phylogenetic tree. The only differences from this general scheme to the specific algorithms are the selection criterion used in the selection step, the branch length formula used in the joining step, and the dissimilarity formula used in the reduction step.

\begin{algorithm}[!ht]
	\textbf{Input:} A distance matrix $D$ over a set of elements $S$.
	
	\textbf{Output:} A phylogenetic tree $T$ over $S$.
	
	\vspace{0.3cm}
	
	\textbf{Initialization:} Initialize the cluster-set $C$ by defining a singleton cluster $C_i = \{i\}$ for every element $i \in S$. Initialize output tree $T = \emptyset$.
	
	\vspace{0.3cm}
	
	\textbf{Loop:} While $|C| > 1$ do:
	\begin{enumerate}
		\item \textbf{Selection:} Select a pair of distinct clusters \{$C_i$, $C_j$\} $\subseteq C$ of minimal dissimilarity under $D$. 
		\item \textbf{Joining:} Remove $C_i,C_j$ from the cluster set $C$ and replace them with $C_u = \{C_i \cup C_j\}$. Calculate the branch length for both elements, namely $D_{iu}$ and $D_{ju}$ and add $C_u$ to the tree $T$.
		\item \textbf{Reduction:} Calculate the dissimilarity $C_{uk}$ for every $C_k \in C' \setminus \{C_i \cup C_j\}$.
	\end{enumerate}
	\textbf{Finalization:} Return the tree $T$.
	\caption{General scheme for hierarchical agglomerative clustering algorithms based on distance matrices.}
	\label{alg:pseudocode}
\end{algorithm}

\subsection{Globally Closest Pairs}

\ac{GCP} based algorithms are widely used in phylogeny. The selection criterion for these algorithms is always the same, that is, choose the smallest pairwise distance, and in case of a tie choose randomly. The branch length formula used in the joining step also never changes and can be defined as $D_{ij}/2$. The dissimilarity formula used in the reduction step is where these algorithms differ.

The algorithm \ac{SL} defines the dissimilarity between two clusters as the minimum value, as show in Equation \ref{eq:sl}.
\begin{equation} \label{eq:sl}
    D_{uk} = min\{D_{ik}, D_{jk}\}
\end{equation}

The algorithm \ac{CL} defines it as the maximum value, as show in Equation \ref{eq:cl}.
\begin{equation} \label{eq:cl}
    D_{uk} = max\{D_{ik}, D_{jk}\}
\end{equation}

The algorithm \ac{UPGMA} defines it as the average dissimilarity, as shown in Equation \ref{eq:upgma}.
\begin{equation} \label{eq:upgma}
    D_{uk} = \frac{|C_i| \cdot D_{ik} + |C_j| \cdot D_{jk}}{|C_i| + |C_j|}
\end{equation}

The algorithm \ac{UPGMC} adjusts the \ac{UPGMA} to the cluster size, as shown in Equation \ref{eq:upgmc}.
\begin{equation} \label{eq:upgmc}
    D_{uk} = \frac{|C_i| \cdot D_{ik} + |C_j| \cdot D_{jk}}{|C_i| + |C_j|} - \frac{|C_i| \cdot |C_j| \cdot D_{ij}}{(|C_i| + |C_j|) ^ 2}
\end{equation}

The algorithm \ac{WPGMA} defines it by weighting the two clusters in half, as shown in Equation \ref{eq:wpgma}.
\begin{equation} \label{eq:wpgma}
    D_{uk} = \frac{D_{ik} + D_{jk}}{2}
\end{equation}

The algorithm \ac{WPGMC} adjusts the \ac{WPGMA} to the cluster size, as shown in Equation \ref{eq:wpgmc}.
\begin{equation} \label{eq:wpgmc}
    D_{uk} = \frac{D_{ik} + D_{jk}}{2} - \frac{D_{ij}}{4}
\end{equation}

\subsection{Neighbour Joining}

The Neighbor Joining algorithm is the most commonly used algorithm in phylogenetics and many variants of this algorithm have been introduced over the years. While some try to optimize the formulas used by \ac{NJ} to better estimate the true and optimal tree, others try to improve its efficiency, both in terms of running time and memory usage. Neighbour-Joining variants differ from one another in all three steps of the general scheme.

This project will have in mind the first \ac{NJ} algorithm, \ac{NJ} by Saitou and Nei \cite{njsn}, its successor, \ac{NJ} by Studier and Keppler \cite{njsk}, and some of its variants, namely \ac{UNJ} \cite{unj}, BioNJ \cite{bionj}, \ac{FNJ} \cite{fnj}, and \ac{RNJ} \cite{rnj}.

\paragraph{Selection}

The selection criterion used in the selection step by Saitou and Nei is expressed by the minimum dissimilarity, given by Equation \ref{eq:selectionsn}. 
\begin{equation}\label{eq:selectionsn}
    Q_{ij} = \frac{D_{ij}}{2} + \frac{\sum\limits_{\substack{k=1}}^{C} (D_{ik} + D_{jk})}{2 \cdot (|C| - 2)}  + \frac{\sum\limits_{\substack{k=1}}^{C} \cdot \sum\limits_{\substack{l=k}}^{C} D_{kl}}{|C| - 2}
\end{equation}

The previous criterion was then simplified into Equation \ref{eq:selectionsk} by Studier and Keppler. \ac{UNJ} also relies on this selection criterion.

\begin{equation}\label{eq:selectionsk}
    Q_{ij} = (|C|-2) \cdot D_{ij} - \sum\limits_{\substack{k=1}}^{C}(D_{ik}+D_{jk})
\end{equation}

Although the implementation by Saitou and Nei is the base algorithm, the implementation by Studier and Keppler is the one that other \ac{NJ} algorithms derive from. That is because the selection criterion used by Studier and Keppler, besides being equal to the one used by Saitou and Nei, also has the advantage of leading to a complexity of $\mathcal{O}(n^3)$ instead of $\mathcal{O}(n^5)$.

\ac{FNJ} algorithm uses a similar criterion as the \ac{NJ} by Studier and Keppler, but instead of choosing the minimum in $Q$ chooses from a different set, called \textit{visible} set, of size $O(n)$ that contains all \textit{visible} pairs. A pair $(C_i, C_j)$ is \textit{visible} if $C_j$ is the minimum, as defined in Equation \ref{eq:selectionfnj}.
\begin{equation}\label{eq:selectionfnj}
    C_j = \min \left\{\sum\limits_{\substack{k=1,\\k \neq i}}^{S} Q_{ik}\right\}
\end{equation}

Unlike \ac{NJ}, that looks for a minimum among all transformed distances, \ac{RNJ} looks for two taxa that have minimal transformed distance between them as compared to their transformed distances to all other taxa.

BioNJ defines a simple selection criterion based on variances of evolutionary distance, that is expressed by the minimum variance of the distance matrix, given by $Qij = D_{ij}/l_s$, where $l_s$ represents the sequence length.

\paragraph{Joining}

The branch length formula used in the joining step by UNJ to calculate $D_{iu}$ and $D_{ju}$ is defined by Equation \ref{eq:branchunj} and $D_{ju} = D_{ij} - D_{iu}$.
\begin{equation} \label{eq:branchunj}
    D_{iu} = \frac{D_{ij}}{2} + \frac{\sum\limits_{\substack{k=1}} ^ {C} |C_k| \cdot (D_{ik} - D_{jk})}{2 \cdot (|C| - |C_u|)}
\end{equation}

All other variants, redefine this formula to Equation \ref{eq:branchnj}, by setting $|C_k|=1$ and replacing $(|C|-|C_u|)$ by $ \sum\limits_{\substack{k=1}}^{C}|C_k|$ which is then equal to $(|C|-2)$.
\begin{equation} \label{eq:branchnj}
    D_{iu} = \frac{D_{ij}}{2} + \frac{\sum\limits_{\substack{k=1}} ^ {C} (D_{ik} - D_{jk})}{2 \cdot (|C| - 2)}
\end{equation}

\paragraph{Reduction}

In the reduction step is defined a general dissimilarity formula expressed by Equation \ref{eq:dissimilarity}, where $D_{iu}$ and $D_{ju}$ are given by the branch length formula, and $\lambda$ is the weight the variant assigns to each branch.
\begin{equation}\label{eq:dissimilarity}
    D_{uk}=\lambda \cdot (D_{ik} - D_{iu}) + (1 - \lambda) \cdot (D_{jk} - D_{ju})
\end{equation}

Studier and Keppler, Saitou and Nei, and \ac{FNJ} define $\lambda = 1/2$, to provide both original branch lengths an equal weight. Saitou and Nei however do not consider the newly computed branch lengths, thus defining them as $D_{iu} = D_{ju} = 0$. Besides joining matrix $D$, \ac{FNJ} also joins the \textit{visible} set by removing the previously selected \textit{visible} pair and adding a new \textit{visible} pair for $C_u$.

\ac{UNJ} defines the weight $\lambda$ as proportional to the number of elements contained in the clusters, as shown in Equation \ref{eq:lambdaunj}, hence giving the same weight to each element.
\begin{equation}\label{eq:lambdaunj}
     \lambda = \frac{|C_i|}{|C_i| + |C_j|}
\end{equation}

BioNJ defines the weight $\lambda$ according to the variance of the element, as defined in Equation \ref{eq:lambdabionj}, where $\lambda \in [0,1]$.
\begin{equation}\label{eq:lambdabionj}
    \lambda = \frac{1}{2} + \frac{\sum\limits_{\substack{k=1,\\k \neq i,j}}^{C} (D_{jk} - D_{ik})}{2 \cdot (|C| - 2) \cdot D_{ij}}
\end{equation}

\subsection{Minimum Spanning Tree}

Minimum Spanning Tree algorithms were developed following a graph theoretic approach. Therefore, the problem of determining the minimal phylogenetic tree will be discussed regarding graph theory. A phylogenetic tree is a graph that is connected but does not contain any cycles. A graph is said to be connected if there exists at least one path between every pair of distinct points.

These algorithms have the particularity of skipping the reduction step as there is no need to update the overall pairwise distances, and thus also not needing the branch length formula in the joining step. There is a variety of algorithms based on this, however this project will focus itself on \ac{goeBURST} \cite{goeburst} and GrapeTree. 

\paragraph{Globally Optimized eBURST}

The \ac{goeBURST} algorithm is a globally optimized implementation of the \ac{eBURST} algorithm that identifies alternative patterns of descent for several bacterial species, using the algorithm by Kruskal \cite{kruskal}. It implements the simplest model for the emergence of clonal complexes, where a given sequence increases in frequency in the population, as a consequence of a fitness advantage or of random genetic drift, becoming a founder clone in the population. This increase is accompanied by a gradual diversification of that sequence, by mutation and recombination, forming a cluster of phylogenetically closely related strains.

The diversification of the “founding” sequence is reflected in the appearance of \ac{ST} differing only in one housekeeping gene sequence from the founder sequence – \ac{SLV}. Further diversification of those \ac{SLV}s will result in the appearance of variations of the original sequence with more than one difference in the allelic profile: \ac{DLV}, \ac{TLV}, and so on.

This algorithm defines its selection criterion as the smallest pairwise distance, and in case of a tie chooses according to the highest number of \ac{SLV}s, \ac{DLV}s, \ac{TLV}s, occurrence frequency, and then according to the lowest id. There is a variant of \ac{goeBURST}, named \ac{goeBURST} Full \ac{MST}, that extends the previously defined rule up to $n$LV level, where $n$ is equal to the number of loci in a strain. If $n$ is defined as one, two or three, the results of this algorithm will be equivalent to the results of \ac{goeBURST} at the levels of \ac{SLV}, \ac{DLV} and \ac{TLV} respectively.

These two algorithms can also be performed dynamically, that is, they can be applied to an already built phylogenetic tree by adding new sequences, instead of having to compute all sequences again to build the phylogenetic tree. However, this project will focus itself only on the static versions.

\paragraph{GrapeTree}

GrapeTree is a novel \ac{MST} algorithm that is better suited for handling missing data than classical \ac{MST} algorithms. It uses the algorithm by Edmonds \cite{edmonds} and a directional measure based on normalized asymmetric Hamming-like distances, to compute a directed minimum spanning tree.

The \ac{eBURST} approach presumes that a clonal complex (lineage) is founded by a founder genotype and that genetic variants of that founder reflect the progressive accumulation of additional variations over time. A further implicit belief is that the number of variants decreases with distance from the founder genotype, such that the founder is equated with the central genotype with the greatest number of single locus variants, and edges between nodes are ordered based on their allelic distances. In case of a tie for directionality of connections, the founder status is assigned to the node with the greater number of \ac{SLV}s, \ac{DLV}s, \ac{TLV}s, and/or number of strains assigned to that \ac{ST}.

At the levels of resolution of core genes that are present in most isolates of a species, \ac{cgMLST}, the founder genotype may not be present in a comparison, which renders the \ac{eBURST} model inappropriate for tie-breaking. Instead of depending on the preconceived properties of a theoretical founder genotype, GrapeTree simply chooses central nodes between multiple co-optimal branches on the basis of the harmonic mean of allelic distances. A central node is defined as the genotype for any given population that has the smallest average allelic distance to all other genotypes in the same population. The selection criterion for the GrapeTree algorithm is defined as the smallest pairwise distance. However in case of a tie, it is defined as the minimum harmonic mean of the allelic distances rather than an arithmetic mean, to give higher weights to sequences with smaller allelic distances to other sequences. Equation \ref{eq:selectiongrapetree} showcases this harmonic mean.
\begin{equation}\label{eq:selectiongrapetree}
    Q_i = \frac{|C| - 1}{\sum\limits_{\substack{j=1,\\j \neq i}}^{C} D_{ij} ^ {-1}}
\end{equation}

Edmonds algorithm \cite{edmonds} is used as the base \ac{MST} algorithm to attempt to minimize the sum of edge lengths in the tree. However, the resulting tree does not necessarily represent true phylogenetic relationships between sequences, because allelic distances do not always correlate with divergence time. Therefore, a \ac{LBR} optimization is subsequently implemented to account for these discrepancies. This local optimization will be further discussed in the next section.
\section{Optimization}

The purpose of a local optimization is to minimize the total weight of the given phylogenetic tree. Every clustering algorithm that uses a dissimilarity formula, which is both convex and commutative, can be locally optimized. A dissimilarity formula is convex if the distance between any cluster $C_k$ to the new cluster $C_u$, where $C_u = C_i \cup C_j$, lies between the distance from that cluster $C_k$ to $C_i$ and $C_j$. And it is said to be commutative if given four arbitrary clusters $\left\{C1; C2; C3; C4\right\}$, the dissimilarity matrix obtained by first joining $\left\{C1; C2\right\}$ and then joining $\left\{C3; C4\right\}$ is equal to the dissimilarity matrix obtained by first joining $\left\{C3; C4\right\}$ and then joining $\left\{C1; C2\right\}$.

Excluding \ac{NJ} and variants, all of the phylogenetic algorithms mentioned before have a commutative and convex dissimilarity formula. \ac{NJ} algorithms cannot assure the convexity property, because their dissimilarity formula depends on the new branch lengths calculated for the two joined elements, and the formula used to compute it does not guarantee convexity, thus leading to possible negative branch lengths. Although the dissimilarity formula in \ac{MST} algorithms is nonexistent, it is still considered as convex and commutative. However, because these algorithms already result in the phylogenetic tree with the minimum distances, applying any local optimization to these trees will result in a tree of equal weight.

\paragraph{Generalization}

Local optimizations tend to follow a general scheme, in which they differ in the selection criterion used to select the next edge to substitute, and the joining criterion used to select the new edges of the phylogenetic tree. This general scheme is portrayed in Algorithm \ref{alg:optimizationpseudocode}.

\begin{algorithm}[!ht]
	\textbf{Input:} A phylogenetic tree $T$ over a set of elements $S$.
	
	\textbf{Output:} The phylogenetic tree $T$.
	
	\vspace{0.3cm}
	
	\textbf{Initialization:} Initialize the set $E$ with the edges of the tree $T$. 
	
	\vspace{0.3cm}
	
	\textbf{Loop:} While $|E| > 0$ do:
	\begin{enumerate}
		\item \textbf{Selection:} Select an edge $(u \rightarrow v)$ of the set $E$ and remove it from the tree $T$, dividing it into two sub-trees $T_u$ (containing $u$) and $T_v$ (containing $v$).
		\item \textbf{Joining:} Find two vertices $w$ and $z$ that best connect the two sub-trees by an edge $(w \rightarrow z)$.
		\item \textbf{Reduction:} Remove the edge $(u \rightarrow v)$ from $E$ and add the edge $(w \rightarrow z)$ to $T$.
	\end{enumerate}
	\textbf{Finalization:} Return the tree $T$.
	\caption{General scheme for local optimization algorithms.}
	\label{alg:optimizationpseudocode}
\end{algorithm}

\paragraph{Local Branch Recrafting}

Despite the fact that \ac{MST} algorithms already result in the phylogenetic tree with the minimum distances, the resulting tree may not necessarily represent true phylogenetic relationships between strains. That may happen as a result of allelic distances not always correlating with divergence time. For that reason, the GrapeTree algorithm still applies a local optimization over it, named \ac{LBR}, which depends on the likelihoods of a contemporary model versus an ancestor-descendent model. Its joining criterion consists of finding two nodes that have the minimum harmonic distance if the contemporary model has a higher or equal likelihood to the ancestor-descendent model. Or, otherwise that have the minimum dissimilarity in relation to $u$ and $v$ respectively. Its selection criterion consists of selecting the shortest edge of the set, and there is an additional step in the reduction to add the edge chosen by the joining criterion to the set $E$, if it does not have the minimum dissimilarity between all edges of the tree.

\paragraph{Rearrangement Measures}

Local optimizations can be based on transformation processes usually found in tree comparison measures. Such measures are seldom used in practice for large studies, as they are expensive to calculate if the trees are dissimilar. However, in the context of this project, they will be useful as they provide multiple alternatives to transforming a tree. These rearrangement measures include \ac{SPR} \cite{spr}, \ac{NNI} \cite{nni}, and \ac{TBR} \cite{tbr}. All of these share the same selection criterion that consists of selecting a random edge of the set. The \ac{SPR} measure defines its joining criterion as the selection of the $u$ vertex, and a new vertex $w$ resulting from the subdivision of an edge of $T_v$ and the suppression of the vertex $v$ from the tree. The \ac{NNI} measure is equivalent to the \ac{SPR} where the vertex $v$ and the new vertex $w$ share a neighbour. The \ac{TBR} measure is similar to the \ac{SPR}, with the exception that it joins instead of the vertex $u$, a new vertex $z$ resulting from the subdivision of an edge of $T_u$ and the suppression of the vertex $u$ from the tree.

\section{Related Work}

There are hundreds of computational phylogenetics tools out there that are commonly used in comparative genomics, cladistics, and bioinformatics. Although they all try to achieve the same goal, which is to build a phylogenetic tree, they all differ widely in multiple aspects.

These tools may represent phylogenetic trees in different formats, such as Newick, Nexus, or even a format of their own. They may even deal with different input formats, such as FASTA, \ac{SNP}, \ac{MLST}, \ac{MLVA}, among others. They may also have different implementations, producing different phylogenetic trees by relying on a set of algorithms for estimating phylogenies, such as Neighbour Joining, Maximum Parsimony, Globally Closest Pairs, Bayesian phylogenetic inference, and Maximum Likelihood. They can also rely on several different distance calculation and correction formulas, such as Hamming, Jukes-Cantor, Kimura, and so on. Some tools may provide local optimization algorithms, such as \ac{LBR}, \ac{SPR}, \ac{NNI}, and \ac{TBR}, while others may not provide any.

These tools can have two different purposes. Some are meant to be used by other tools, in the format of a library or command-line application. While others are created to be used directly by the final user, as a desktop or web application, which may be free or paid. They can be implemented in different languages and used in different platforms or even be specific to one platform. However, besides differing in many aspects, these tools may also share some unappealing aspects, like not being easily integrated into existing phylogenetic analysis workflows, not supporting a common API between algorithms, and not always providing efficient implementations.

Some examples of the most well-known libraries are PHYLIP \cite{phylip}, PhyML \cite{phyml}, RAxML \cite{raxml}, PAUP* \cite{paup*}, MrBayes \cite{mrbayes} and MEGA \cite{mega}. And, some examples of the most frequently used desktop and web applications, that are free, are PHYLOViZ and PHYLOViZ Online. The major differences between these tools, regarding the phylogenetic analysis workflow for distance matrix based algorithms, can be seen in Table \ref{table:libraries}. In this table, the first and last columns represent the input and output formats supported by each tool, the columns with a number between parentheses represent the four steps of the phylogenetic analysis workflow discussed in this chapter, and the second to last column refers the other four columns and represents the steps of the workflow that provide an output file.

\afterpage{
    \begin{landscape}
        \begin{table}
            \centering
            \begin{tabular}{ | c | c | c | c | c | c | c | c | }
                \hline
                    \diagbox{Tool}{Feature} & \begin{tabular}{@{}c@{}}Input \\ Format\end{tabular} & \begin{tabular}{@{}c@{}}Distance \\ Calculation (1)\end{tabular} & \begin{tabular}{@{}c@{}}Distance \\ Correction (2)\end{tabular} & \begin{tabular}{@{}c@{}}Inference \\ Algorithm (3)\end{tabular} & \begin{tabular}{@{}c@{}}Local \\ Optimization (4)\end{tabular} & \begin{tabular}{@{}c@{}}Output \\ Processing\end{tabular} & \begin{tabular}{@{}c@{}}Output \\ Format\end{tabular} \\
                \hline
                    PHYLIP & PHYLIP & Hamming & Fitch-Margoliash & \begin{tabular}{@{}c@{}}\ac{UPGMA} \\ \ac{NJ} by Saitou \& Nei\end{tabular} & Robinson-Foulds & \begin{tabular}{@{}c@{}}(1) \\ (2) \\ (3) \\ (4)\end{tabular} & Newick \\
                \hline
                    PhyML & \begin{tabular}{@{}c@{}}PHYLIP \\ Nexus\end{tabular} & Hamming & \begin{tabular}{@{}c@{}}JC69 \\ K80 \\ F81 \\ F84 \\ HKY85 \\ TN93 \\ GTR\end{tabular} & BioNJ & \begin{tabular}{@{}c@{}}\ac{NNI} \\ \ac{SPR}\end{tabular} & \begin{tabular}{@{}c@{}}(3) \\ (4)\end{tabular} & Newick \\
                \hline
                    RAxML & \begin{tabular}{@{}c@{}}PHYLIP \\ FASTA\end{tabular} & Hamming & \begin{tabular}{@{}c@{}}JC69 \\ K80\end{tabular} & - & \ac{NNI} & \begin{tabular}{@{}c@{}}(3) \\ (4)\end{tabular} & Newick \\
                \hline
                    PAUP* & Nexus & Hamming & HKY85 & \begin{tabular}{@{}c@{}}\ac{UPGMA} \\ \ac{NJ} by Saitou \& Nei \\ \end{tabular} & - & (3) & Nexus \\
                \hline
                    MrBayes & Nexus & Hamming & \begin{tabular}{@{}c@{}}JC69 \\ K80 \\ F81 \\ HKY85 \\ GTR\end{tabular} & \begin{tabular}{@{}c@{}}BioNJ \\ \ac{NJ} by Saitou \& Nei\end{tabular} & - & (3) & Nexus \\
                \hline
                    MEGA & MEGA & Hamming & \begin{tabular}{@{}c@{}}Jukes-Cantor \\ Tajima-Nei \\ Kimura 2-Parameter \\ Tamura 3-Parameter \\ Tamura-Nei \\ Log-Det\end{tabular} & \begin{tabular}{@{}c@{}}\ac{UPGMA} \\ \ac{NJ} by Saitou \& Nei\end{tabular} & - & (3) & Newick \\
                \hline
                    PHYLOViZ & \begin{tabular}{@{}c@{}}\ac{MLST} \\ \ac{MLVA} \\ \ac{SNP}\end{tabular} & Hamming & - & \begin{tabular}{@{}c@{}}\ac{goeBURST} \\ \ac{CL} \\ \ac{SL} \\ \ac{UPGMA} \\ \ac{WPGMA} \\ \ac{NJ} by Saitou \& Nei \\ \ac{NJ} by Studier \& Keppler\end{tabular} & - & (3) & Custom JSON \\
                \hline
                    \begin{tabular}{@{}c@{}}PHYLOViZ \\ Online\end{tabular} & \begin{tabular}{@{}c@{}}\ac{MLST} \\ \ac{MLVA} \\ FASTA \\ Newick\end{tabular} & Hamming & - & \ac{goeBURST} & - & (3) & Custom JSON \\
                \hline
            \end{tabular}
            \caption{Major differences between some of the most well-know phylogenetic tools, regarding the phylogenetic analysis workflow.}
            \label{table:libraries}
        \end{table}
    \end{landscape}
}

\section{Discussion}

The phylogenetic analysis workflow can be summarized into four consecutive steps, the distance calculation, distance correction, inference algorithm, and local optimization steps. The first step consists of producing a distance matrix from a dataset, including several sequences, through a distance calculation method, such as Hamming, GrapeTree, or Kimura, that calculates the distances between each pair of sequences of the dataset. The dataset can be represented in several formats, including \ac{MLST}, \ac{MLVA}, FASTA, and \ac{SNP}. The second step takes a distance matrix and corrects each distance using a correction formula, such as Jukes-Cantor. This step is optional, thus it may be skipped. The third step transforms a distance matrix into a phylogenetic tree by running a clustering algorithm, such as \ac{goeBURST}, GrapeTree, \ac{UPGMA}, and \ac{NJ} by Studier and Keppler. The phylogenetic tree can be represented in several formats, including Newick and Nexus. And the fourth step takes a phylogenetic tree and tries to locally optimize it through a local optimization algorithm, such as \ac{LBR}. This step is also optional, thus it may be skipped, however it may also be applied several times.

Despite there being many libraries and applications currently available and dedicated to the implementation of the phylogenetic analysis workflow, they all differ widely in many aspects and were not created with integration and extensibility in mind, but instead for a specific purpose. For that reason, it would be beneficial to have a tool that is capable of easily integrating other formats and algorithms into just one common place.
\chapter{Proposed Solution}

This chapter depicts the functional and non functional requirements of the proposed solution, as well as its use cases. It also explains the architecture for the proposed solution and the technologies to be used.

\section{Requirements}

\begin{figure}[!ht]
    \centering
    \includegraphics[scale=0.7]{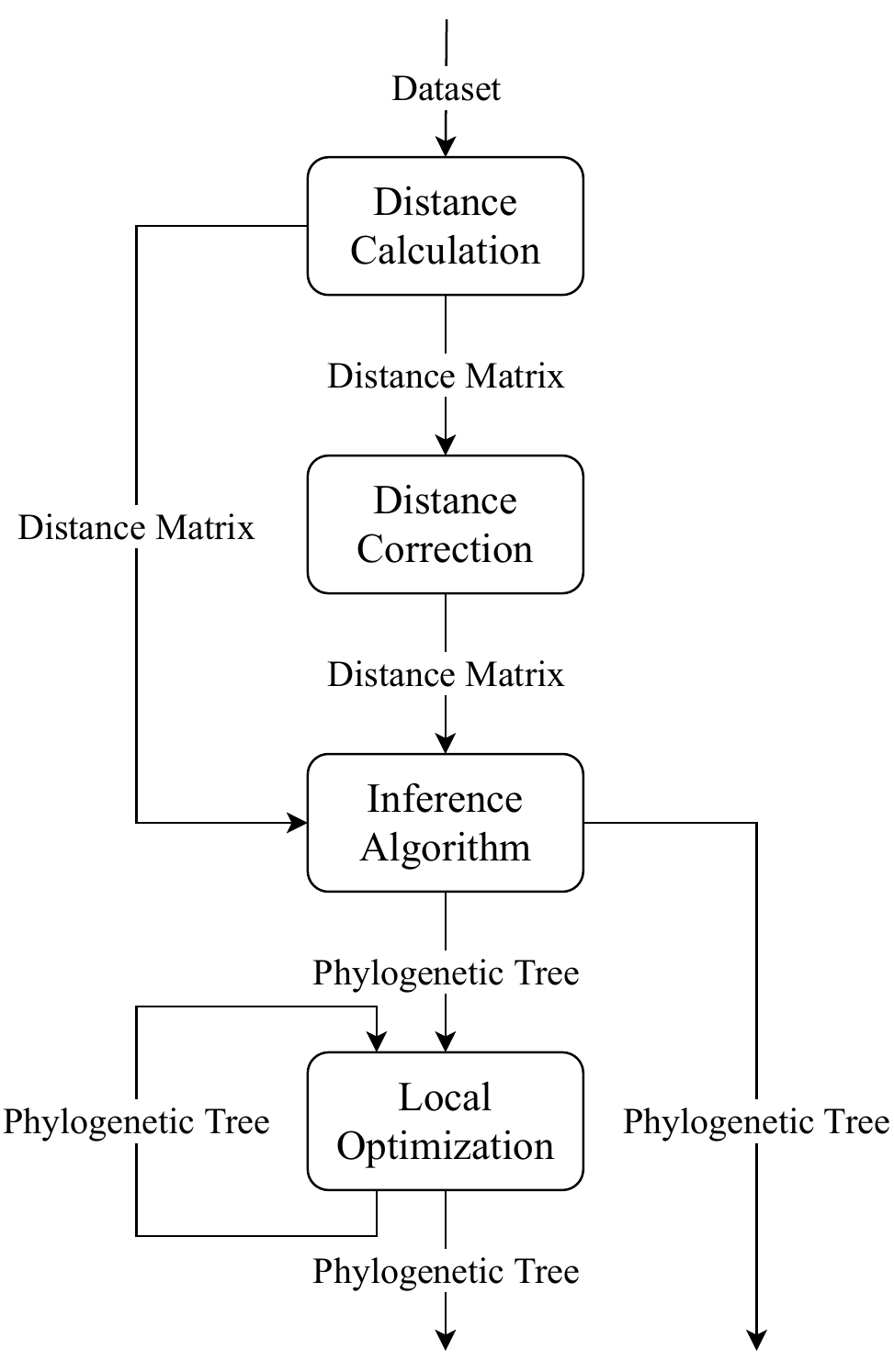}
    \caption{Phylogenetic analysis workflow.}
    \label{fig:workflow}
\end{figure}

The proposed solution for this project revolves around the development of a command line application, titled PhyloLib, that obeys to several functional and non functional requirements, but whose overall requirement is to enable the functioning of the phylogenetic analysis workflow represented in Figure \ref{fig:workflow}. 

Ultimately, this solution should provide implementations for some of the concepts mentioned in the previous chapter, such as some of the existing phylogenetic and optimization algorithms, distance calculation and correction formulas, and dataset, matrix and tree formats.

\subsection{Functional}

In terms of functional requirements, to comply with the previous workflow, this project should:

\begin{enumerate}
    \item Support reading from a file datasets in different formats, such as \ac{MLST}, \ac{MLVA}, \ac{SNP}, and FASTA.
    \item Provide different distance calculation methods to produce a distance matrix, such as Hamming, GrapeTree and Kimura.
    \item Support reading from and writing to a file distance matrices in different formats, such as symmetric and asymmetric.
    \item Allow rectifying a distance matrix with different distance correction methods, such as Jukes-Cantor.
    \item Provide different phylogenetic inference algorithms based on distance matrices to produce a phylogenetic tree, such as \ac{SL}, \ac{CL}, \ac{UPGMA}, \ac{UPGMC}, \ac{WPGMA}, \ac{WPGMC}, \ac{goeBURST}, Edmonds, \ac{NJ} by Saitou and Nei, \ac{NJ} by Studier and Keppler, and \ac{UNJ}.
    \item Support reading from and writing to a file phylogenetic trees in different formats, such as Newick and Nexus.
    \item Allow optimizing a phylogenetic tree multiple times with different local optimization algorithms, such as \ac{LBR}.
    \item Enable executing only specific operations of the workflow, or all of them, at once.
\end{enumerate}

\subsection{Non Functional}

The non functional requirements to have in mind during the development of the project state that the library should:

\begin{enumerate}
    \item Provide high portability and integration, by being able to run in most environments and be integrated with other applications.
    \item Provide high extensibility, by exposing an interface that is easily extensible to more methods, algorithms and formats.
    \item Be highly reusable, by having minimal duplicated code and reusing as much existing code as possible.
\end{enumerate}

\subsection{Use Cases}

Being a command line application, each call to the library should receive its commands through the \ac{CLI} arguments. All commands should be case insensitive and separated by a colon. Each individual command should be represented by its name, type, and options, separated by a space. Each option should be optional and represented by its name preceded by two dashes, or by a letter preceded by a dash, and followed by an equals sign and corresponding value.

The available commands should be \verb|distance|, \verb|correction|, \verb|algorithm|, and \verb|optimization|, each respectively defining an operation in the workflow. All of these commands should be optional and only be declared at most once, except for \verb|optimization| that should be able to be declared multiple times. Also, the order in which the commands are presented should not matter, except for \verb|optimization| that can be declared multiple times. Lastly, as seen in the workflow in Figure \ref{fig:workflow}, every combination of commands should be possible, except if it includes \verb|optimization| and excludes \verb|algorithm|, while still including \verb|distance| and/or \verb|correction|.

The type and options declared in a command should specify its execution, namely the type should identify the implementation for that command, and the options should specify details for that implementation, for example, if the command is \verb|algorithm| then the type may be \verb|goeburst| and an option may be \verb|--lvs|. Each option should not be declared more than once for a command, and the order in which they are declared should not matter. Besides the custom options that may be used by each command and specific type, such as \verb|--lvs| or \verb|-l|, an option that may be declared for every command is \verb|--out| or \verb|-o|, and it defines the output file for the command. Other options that a command may or may not use, depending on its input needs, are \verb|--dataset|, \verb|--matrix|, and \verb|--tree|, or \verb|-d|, \verb|-m|, and \verb|-t|, each respectively defining a data type in the workflow. These three options define input files for the commands, and should be declared if there is no value for that data type in the current context, that is, if the declared commands, that are previous to this one in the workflow, will not provide a value for that data type. All of these file options, including \verb|--out|, should be represented by a format name followed by a colon and a file location.

\begin{figure}[!ht]
    \centering
    \includegraphics[scale=0.7]{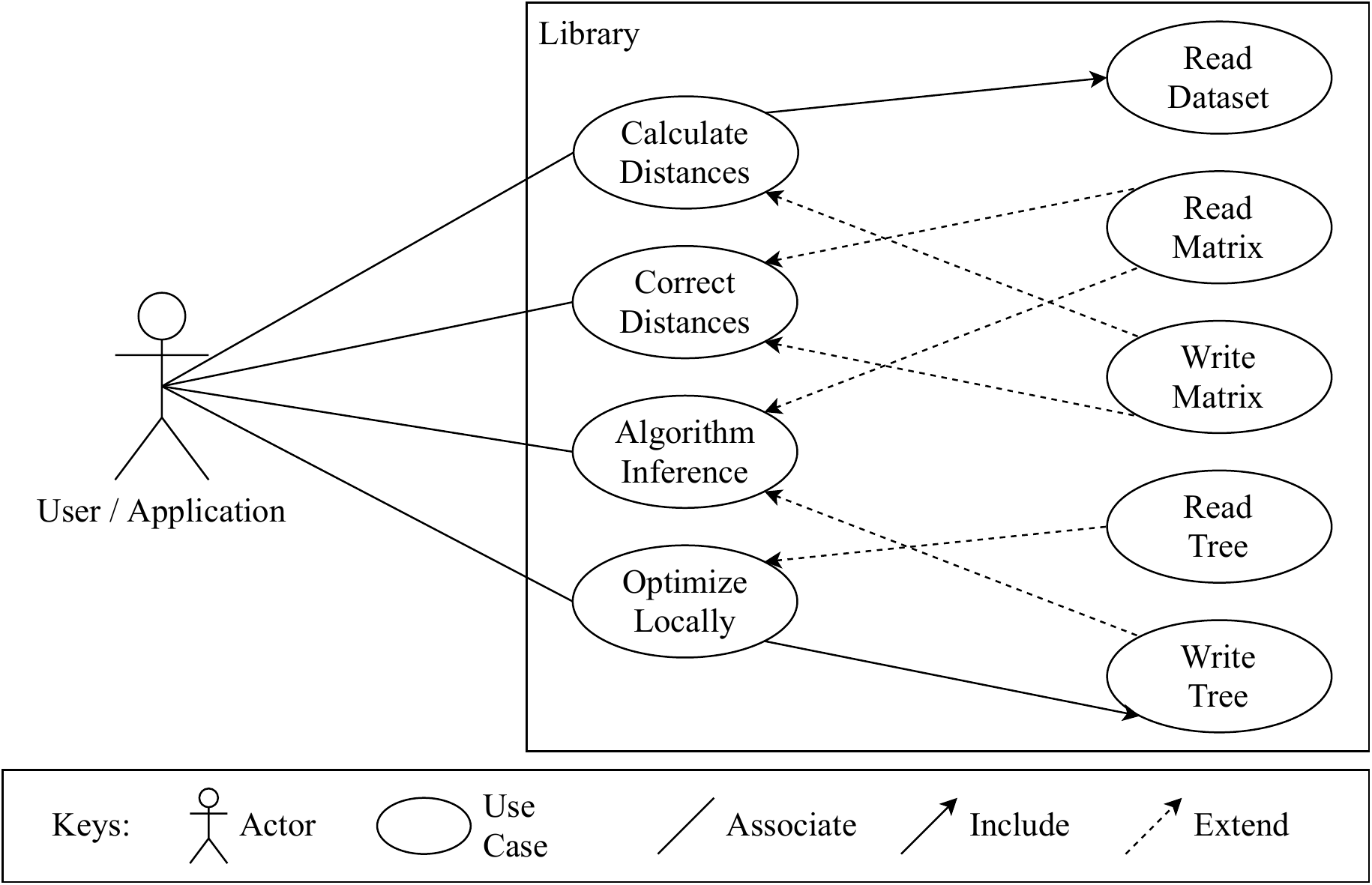}
    \caption{Use cases of the project.}
    \label{fig:usecases}
\end{figure}

This project should include a separate use case for each operation in the workflow, as demonstrated in Figure \ref{fig:usecases}. In this figure, an include relation means the included use case is not optional, and therefore is also applied when the base use case is applied. And, an extend relation means the extending use case might also be applied when the extended use case is applied, making it optional. In this case, the extend relations are applied if an input, in case of a read use case, or an output, in case of a write use case, is specified.

An example use case is to compute the \ac{goeBURST} algorithm with TLVs and the output in Newick to a file \verb|tree.txt|, using the Hamming distance, with the dataset as input in \ac{SNP} format from a file \verb|dataset.txt|, and the \ac{LBR} optimization, with the output in Newick to a file \verb|out.txt|, as follows:
\begin{verbatim}
                phylolib algorithm goeburst --lvs=3 --out=newick:tree.txt :
    distance hamming --dataset=snp:dataset.txt : optimization lbr --out=newick:out.txt
\end{verbatim}

\section{Architecture}

\begin{figure}[!ht]
    \centering
    \includegraphics[scale=0.7]{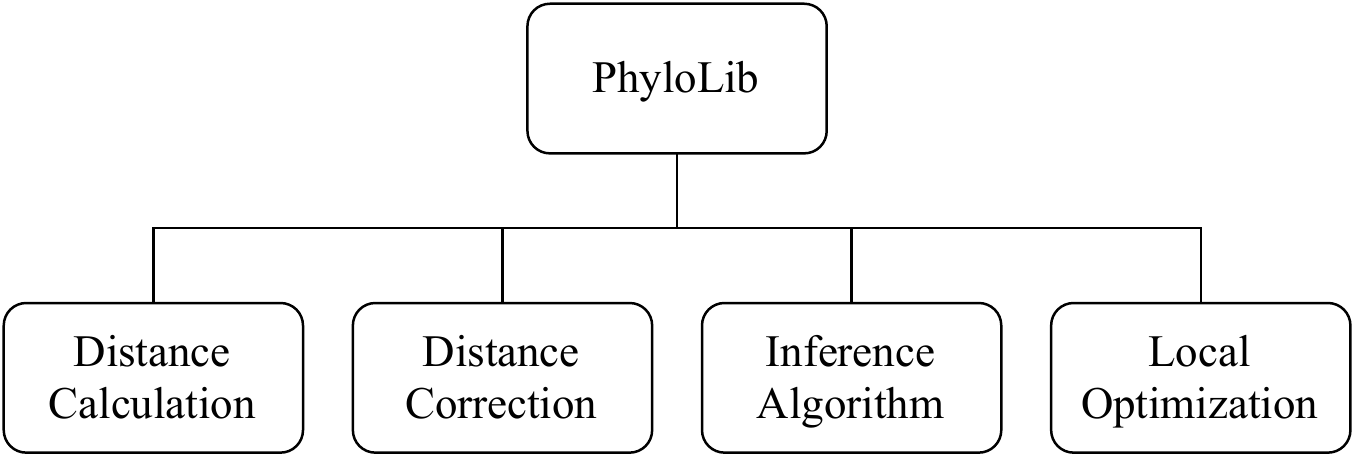}
    \caption{Decomposition view of the main architecture.}
    \label{fig:decomposition}
\end{figure}

The architecture of this project will be decomposed into four operations, the distance calculation, distance correction, inference algorithm, and local optimization, as shown in the decomposition view of the main architecture in Figure \ref{fig:decomposition}.

\begin{figure}[!ht]
    \centering
    \includegraphics[scale=0.7]{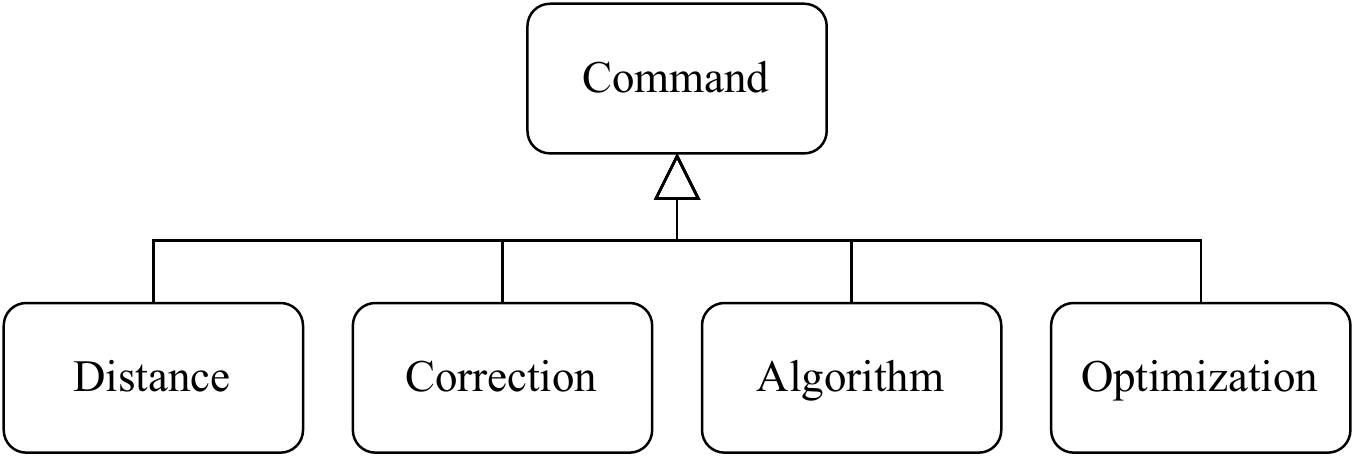}
    \caption{Generalization view of the main architecture.}
    \label{fig:generalization}
\end{figure}

These four operations will be used in accordance with the workflow established in the Requirements section. And they will be translated respectively into four components, the Distance, Correction, Algorithm, and Optimization, as shown in the generalization view of the main architecture in Figure \ref{fig:generalization}. The use of these components will be defined by the user, through the declaration of the commands \verb|distance|, \verb|correction|, \verb|algorithm|, and \verb|optimization|, respectively.

\begin{figure}[!ht]
    \centering
    \includegraphics[scale=0.7]{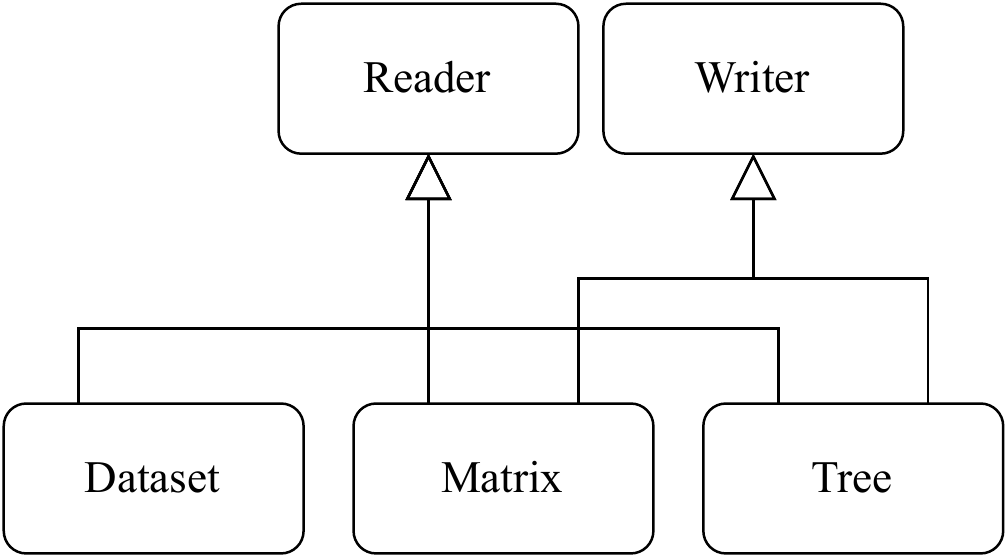}
    \caption{Generalization view of the reading and writing architecture.}
    \label{fig:io}
\end{figure}

This architecture will also be decomposed into two additional operations, the input reading and the output writing, to separate those responsibilities from the previous four operations. They will be translated into two components, the Reader and Writer, and they will be able to define the parsing for the different data types. These data types will be specified by each implementation of the Reader and Writer, namely Dataset, Matrix, and Tree, as shown in the generalization view of the reading and writing in Figure \ref{fig:io}. All of these components implement the Reader and Writer, except for the Dataset, that only implements the Reader, because there is no output of type dataset in the workflow. The use of these components will be defined by the user, through the declaration of the options \verb|--out|, \verb|--dataset|, \verb|--matrix|, or \verb|--tree| in each of the previously mentioned commands.

\begin{figure}[!ht]
    \centering
    \includegraphics[scale=0.7]{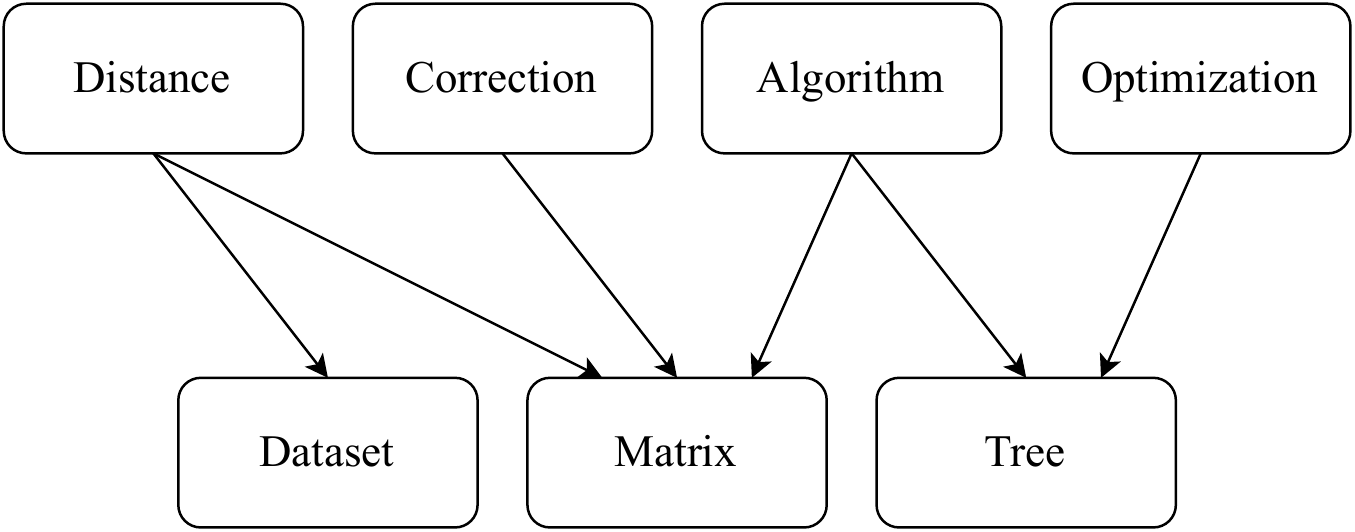}
    \caption{Uses view of the architecture.}
    \label{fig:uses}
\end{figure}

The Dataset, Matrix, and Tree components will be called internally by the Distance, Correction, Algorithm, and Optimization components, in accordance with the uses view of the architecture in Figure \ref{fig:uses}. Note that, this figure only represents the input and output operations that can be observed in the workflow, however each specific type of command may require additional inputs thus requiring uses relations that may not be visible in this figure.

\subsection{Distance Calculation}

\begin{figure}[!ht]
    \centering
    \includegraphics[scale=0.7]{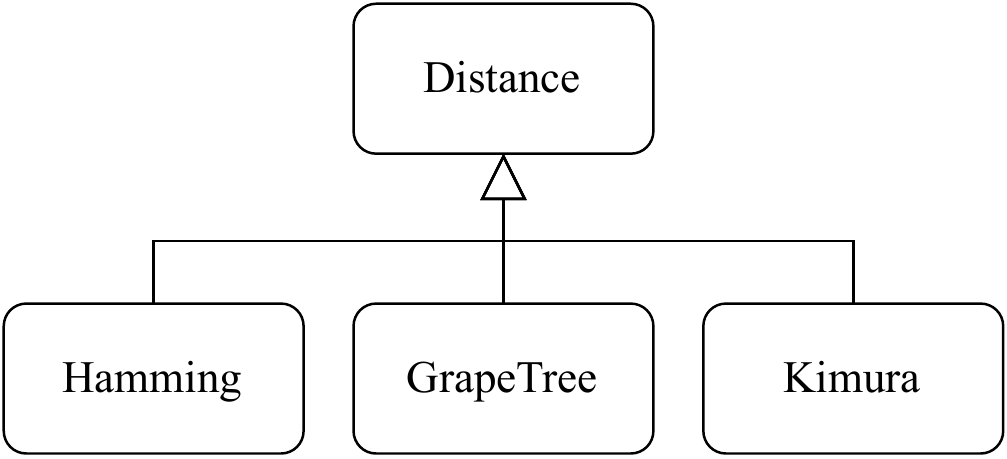}
    \caption{Generalization view of the distance calculation component.}
    \label{fig:distance}
\end{figure}

The Distance component will be responsible for calculating the distances between profiles in a dataset into a distance matrix, based on a specific distance calculation method. This distance calculation method will be specified by each implementation of the Distance component, namely Hamming, GrapeTree, and Kimura, as shown in the generalization view of the Distance component in Figure \ref{fig:distance}. The use of these implementations will be defined by the user, through the definition of the type for the command \verb|distance|, which can be respectively \verb|hamming|, \verb|grapetree|, or \verb|kimura|.

\subsection{Distance Correction}

\begin{figure}[!ht]
    \centering
    \includegraphics[scale=0.7]{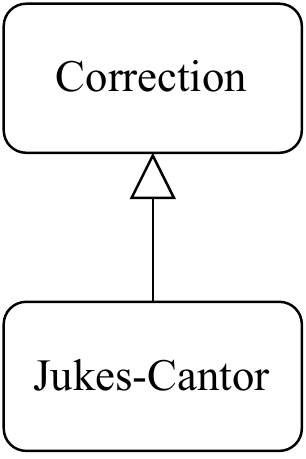}
    \caption{Generalization view of the distance correction component.}
    \label{fig:correction}
\end{figure}

The Correction component will be responsible for rectifying a distance matrix, based on a specific distance correction formula. This distance correction formula will be specified by each implementation of the Correction component, namely Jukes-Cantor, as shown in the generalization view of the Correction component in Figure \ref{fig:correction}. The use of these implementations will be defined by the user, through the definition of the type for the command \verb|correction|, which can be respectively \verb|jukescantor|.

\subsection{Inference Algorithm}

\begin{figure}[!ht]
    \centering
    \includegraphics[scale=0.7]{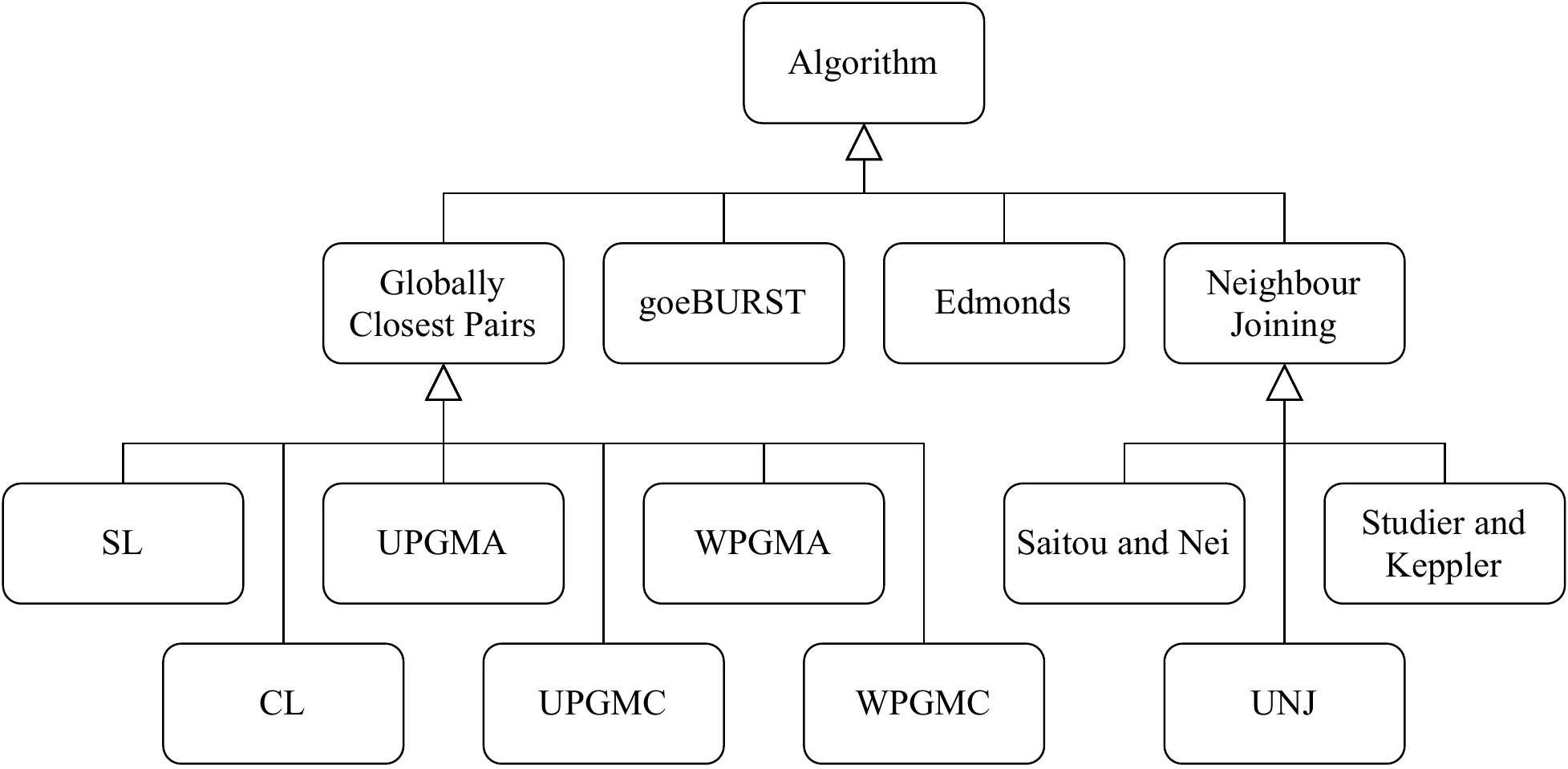}
    \caption{Generalization view of the inference algorithm component.}
    \label{fig:algorithm}
\end{figure}

The Algorithm component will be responsible for processing the distance matrix into a phylogenetic tree, through a specific distance matrix algorithm. This distance matrix algorithm will be specified by each implementation of the Algorithm component, namely \ac{SL}, \ac{CL}, \ac{UPGMA}, \ac{UPGMC}, \ac{WPGMA}, \ac{WPGMC}, \ac{goeBURST}, Edmonds, \ac{NJ} by Saitou and Nei, \ac{NJ} by Studier and Keppler, and \ac{UNJ}, as shown in the generalization view of the Algorithm component in Figure \ref{fig:algorithm}. The \ac{GCP} and \ac{NJ} implementations will be aggregated into categories, specifically Globally Closest Pairs and Neighbour Joining, due to the similarities mentioned in the Background section. The use of these implementations will be defined by the user, through the definition of the type for the command \verb|algorithm|, which can be respectively \verb|sl|, \verb|cl|, \verb|upgma|, \verb|upgmc|, \verb|wpgma|, \verb|wpgmc|, \verb|goeburst|, \verb|edmonds|, \verb|saitounei|, \verb|studierkeppler|, or \verb|unj|.

\subsection{Local Optimization}

\begin{figure}[!ht]
    \centering
    \includegraphics[scale=0.7]{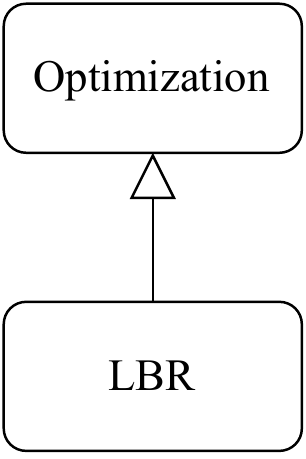}
    \caption{Generalization view of the local optimization component.}
    \label{fig:optimization}
\end{figure}

The Optimization component will be responsible for optimizing the distances in a phylogenetic tree, through a local optimization algorithm. This local optimization algorithm will be specified by each implementation of the Optimization component, namely \ac{LBR}, as shown in the generalization view of the Optimization component in Figure \ref{fig:optimization}. The use of these implementations will be defined by the user, through the definition of the type for the command \verb|optimization|, which can be respectively \verb|lbr|.

\subsection{Dataset Parsing}

\begin{figure}[!ht]
    \centering
    \includegraphics[scale=0.7]{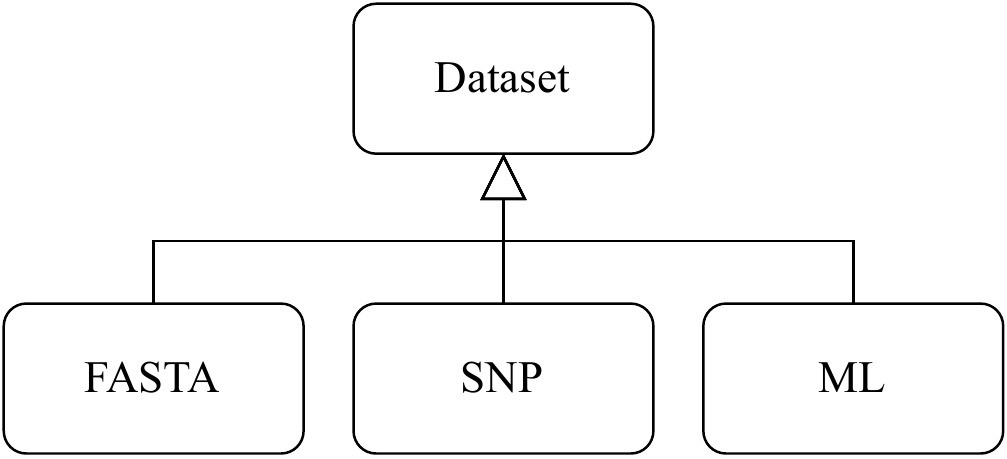}
    \caption{Generalization view of the dataset parsing component.}
    \label{fig:dataset}
\end{figure}

The Dataset component will be responsible for reading a dataset from a specific file location in a specific format. This format will be specified by each implementation of the Dataset component, namely FASTA, \ac{SNP}, and ML, as shown in the generalization view of the Dataset component in Figure \ref{fig:dataset}, where the ML implementation corresponds to the \ac{MLVA} and \ac{MLST} formats. The use of these implementations will be defined by the user, through the definition of the format for the file options \verb|--dataset| and \verb|--out|, which can be respectively \verb|fasta|, \verb|snp|, or \verb|ml|.

\subsection{Distance Matrix Parsing}

\begin{figure}[!ht]
    \centering
    \includegraphics[scale=0.7]{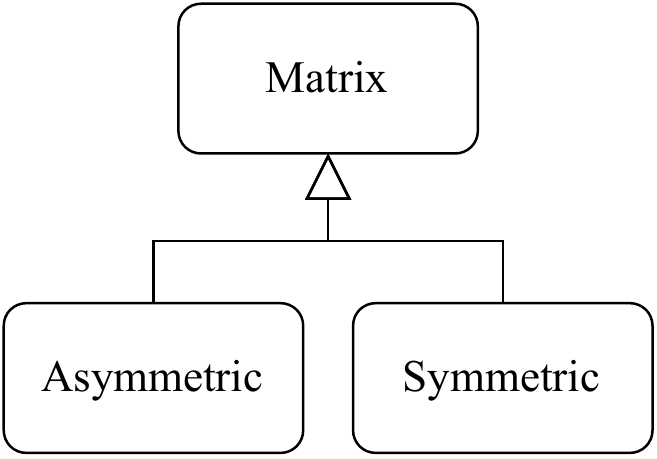}
    \caption{Generalization view of the distance matrix parsing component.}
    \label{fig:matrixparsing}
\end{figure}

The Matrix component will be responsible for reading and writing a distance matrix from and to a specific file location in a specific format. This format will be specified by each implementation of the Matrix component, namely Symmetric, and Asymmetric, as shown in the generalization view of the Matrix component in Figure \ref{fig:matrixparsing}. The use of these implementations will be defined by the user, through the definition of the format for the file options \verb|--matrix| and \verb|--out|, which can be respectively \verb|symmetric| or \verb|asymmetric|.

\subsection{Phylogenetic Tree Parsing}

\begin{figure}[!ht]
    \centering
    \includegraphics[scale=0.7]{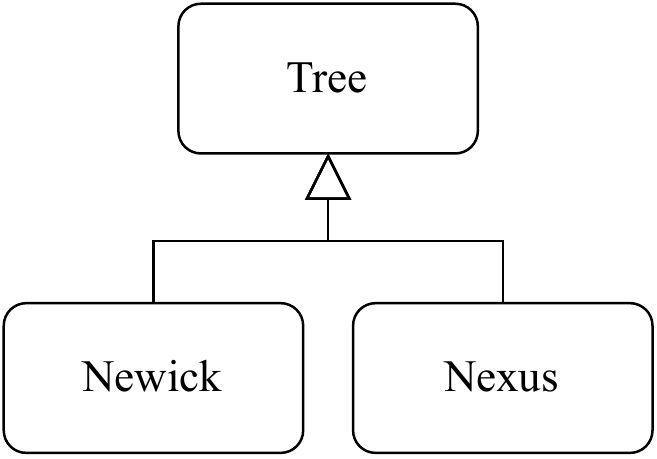}
    \caption{Generalization view of the phylogenetic tree parsing component.}
    \label{fig:tree}
\end{figure}

The Tree component will be responsible for reading and writing a phylogenetic tree from and to a specific file location in a specific format. This format will be specified by each implementation of the Tree component, namely Newick, and Nexus, as shown in the generalization view of the Tree component in Figure \ref{fig:tree}. The use of these implementations will be defined by the user, through the definition of the format for the file options \verb|--tree| and \verb|--out|, which can be respectively \verb|newick| or \verb|nexus|.

\section{Technologies}

The choice of programming language for this project is heavily influenced by the fact that most algorithms are already implemented in Java and JavaScript. However, the choice ends up being Java, due to its portability that enables it to be run in most platforms, as well as its performance capabilities regarding parallelization and multi-threading that JavaScript does not possess. Java is a set of computer software and specifications, that provides a system for developing application software and deploying it in a cross-platform computing environment. Its syntax borrows heavily from C and C++, but object-oriented features are modeled after Smalltalk and Objective-C. Also, memory management in Java is handled through integrated automatic garbage collection performed by the \ac{JVM}.

\section{Discussion}

The proposed solution for this project boils down to the development of a command line application that conforms to the phylogenetic analysis workflow and is highly performant, extensible, reusable, and portable. It should enable reading datasets, distance matrices, and phylogenetic trees from files, calculating and correcting a distance matrix, inferring and locally optimizing a phylogenetic tree, and writing distance matrices and phylogenetic trees to files.

\chapter{Implementation}

This chapter describes more in depth the implementation details that were required to implement the proposed solution of this project. To help better comprehend the implementation, this chapter also provides the \ac{UML} class diagrams for each package of the project.

The proposed solution was implemented bearing in mind an agile methodology, and is publicly available at \url{https://github.com/Luanab/phylolib} as a library along with its issues, milestones, and Javadoc documentation. Aside from the structure related milestones and issues, each of the milestones can be translated into a command or data type, while each of the issues can be translated into an implementation of that command or data type. For testing purposes, the library is hosted in a server as a Docker image.

\section{Structure}

The starting point and main logic of this project is located in the \verb|Main| class, where the whole phylogenetic analysis workflow is connected. This class is responsible for joining all of the main concerns of the project in one place, namely the \ac{CLI} argument parsing, mapping of arguments to commands and data parsers through reflection, workflow setup, exception handling, and logging. The workflow setup includes the specification of the order of execution of the commands and their respective data parsing components. Every file of this project is located inside the \verb|pt.ist.phylolib| package, which is composed of six main parts, the \verb|cli|, \verb|reflection|, \verb|exception|, \verb|logging|, \verb|data|, and \verb|command| packages.

\subsection{Arguments}

\begin{figure}[!ht]
    \centering
    \includegraphics[scale=0.7]{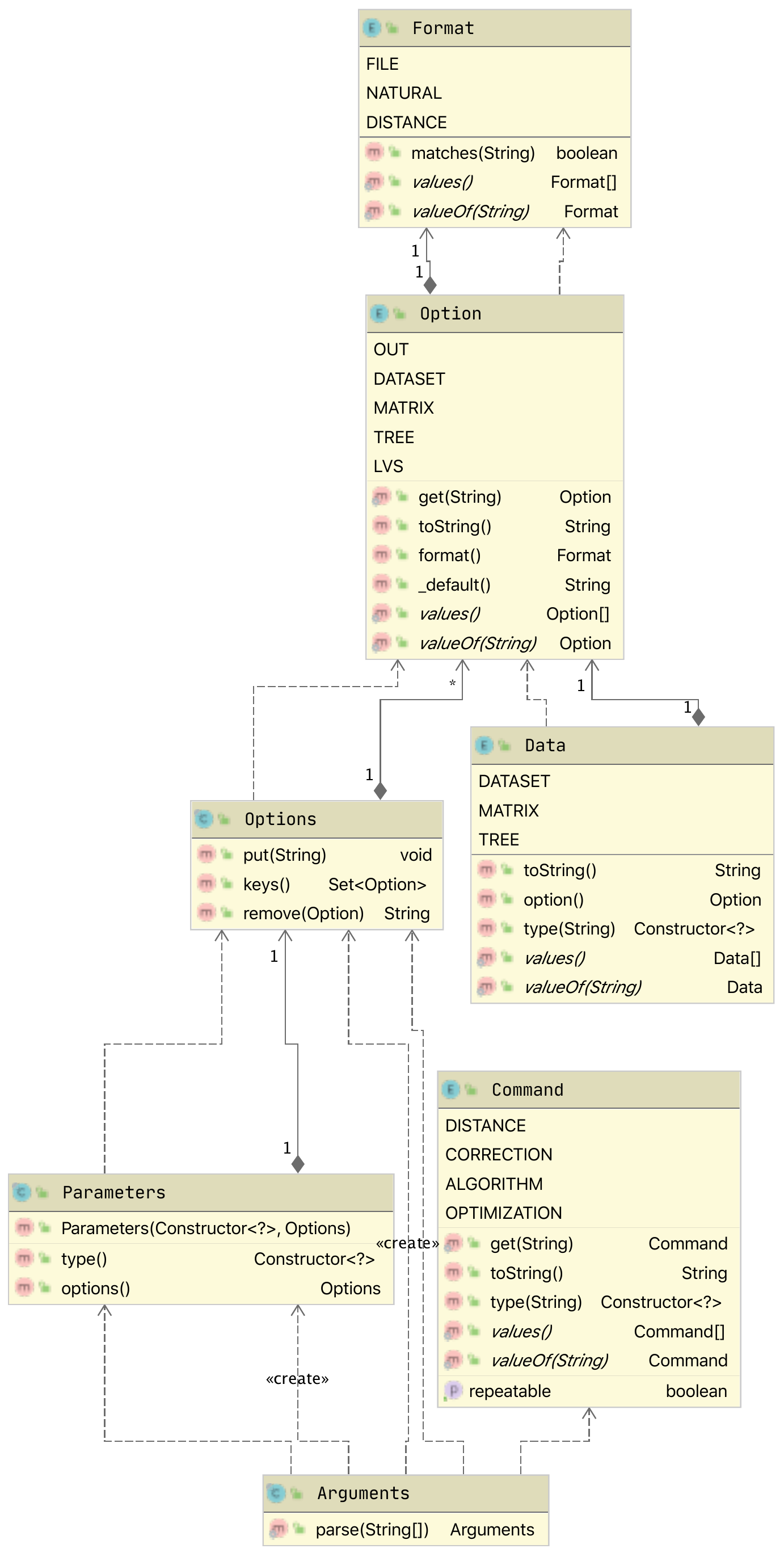}
    \caption{UML class diagram of the cli package.}
    \label{uml:cli}
\end{figure}

The \ac{CLI} argument parsing is focused inside the \verb|Arguments| class in the \verb|cli| package, which decomposes the arguments into commands and parameters. It is called by the \verb|Main| class at the beginning of the workflow to ensure that argument related issues are detected beforehand, that is, before the execution of any command. The \verb|Arguments| class returns to the \verb|Main| class the correspondence between commands and its parameters through a map of objects of the \verb|Command| and \verb|Parameters| classes. The \verb|Command| class is an enum that defines all of the possible commands and their ability to be repeated in an execution of the workflow. The objects of the \verb|Parameters| class serve as the container of the received type and options for a command. The received options themselves are stored in objects of the \verb|Options| class, as a map of objects of the \verb|Option| and \verb|String| classes. The latter contains the value for the option, while the former represents the option itself. The \verb|Option| class is an enum that defines all of the possible options and their corresponding character alias, value format, and optional default value for when the option is not received but is necessary. The value format itself is represented by the \verb|Format| class that is an enum, which defines all of the existing string formats for the options as regular expressions. Lastly, the \verb|Data| class is an enum responsible for connecting each data parser to an input option. The \ac{UML} class diagram that represents this package can be seen in Figure \ref{uml:cli}.

Exceptionally, if the first argument is the command \verb|help|, then the \verb|Arguments| class ceases to continue parsing the arguments and immediately returns to the \verb|Main| class to print the usage message. Otherwise, if no arguments are provided to this library, a received command does not exist or is invalidly repeated, or a command type is not provided or is invalid, then the command parsing stops and an exception is thrown.

\subsection{Reflection}

\begin{figure}[!ht]
    \centering
    \includegraphics[scale=0.3]{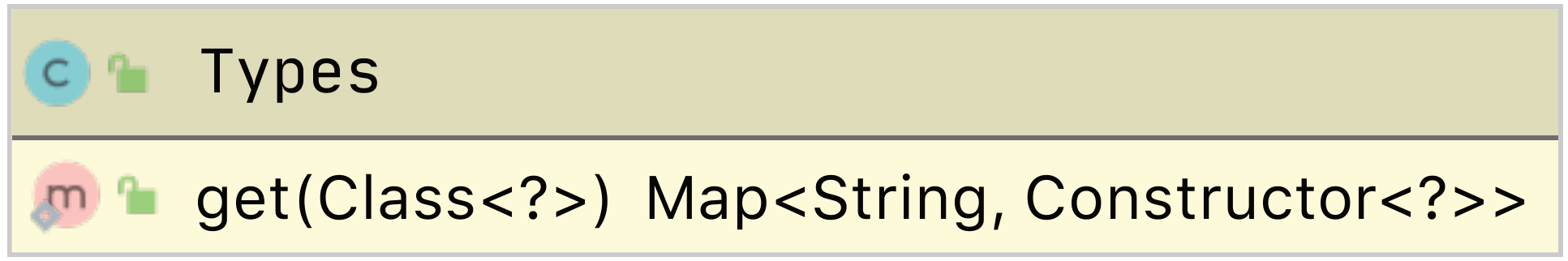}
    \caption{UML class diagram of the reflection package.}
    \label{uml:reflection}
\end{figure}

The mapping of arguments to commands and data parsers is achieved through reflection, with the help of the \verb|Reflections| class from the \verb|org.reflections| external package, and the \verb|Constructor| and \verb|Modifier| classes from the \verb|java.lang.reflect| built-in Java package. This concern is wrapped up in the \verb|Types| class of the package \verb|reflection| of this project. In sum, this class provides a method that retrieves a map with all of the names and constructors of the classes that extend a given class. It only looks inside the \verb|pt.ist.phylolib| package though, to avoid uselessly searching in other dependencies, therefore new implementations must always be defined inside it. The \ac{UML} class diagram that represents this package can be seen in Figure \ref{uml:reflection}.

This functionality is used by the \verb|Command| and \verb|Data| enums from the \verb|cli| package, to dynamically find all of the implementations of the selected commands and data types, and then execute the selected implementation. If the user, for example, tries to execute the \verb|algorithm| command with the type \verb|upgma|, then the \verb|Command| enum will use this functionality to find all classes that extend the \verb|Algorithm| class, and then from these execute the \verb|UPGMA| class.

This approach provides an easy way to extend commands and data types, as it becomes simple to include a new implementation, by either adding it inside the project and compiling again, or just adding it to the classpath, without having to touch any other part of the code. Either way, to execute any implementation of a command or data type it is expected that the name is provided without the package in the command line arguments, so the arguments can be more concise and the user does not have to know about the location of the implementation inside the project. This, however, creates one restriction to all implementations, that is, there should never be any two implementations of the same command or data type with the same name, regardless of the package, as the project will select one of the two implementations randomly.

\subsection{Exceptions}

The exception handling is done at the \verb|Main| class level, where the exceptions are distinguished between user input related and internal issues, and they are logged accordingly. The reason for this is that user input related issues are caused by some faulty user input and thus can be solved by the user alone. While the internal issues are more complex and were not accounted for in the development of this project, and thus should be solved by the developer and not the user itself.

\begin{figure}[!ht]
    \centering
    \includegraphics[scale=0.6]{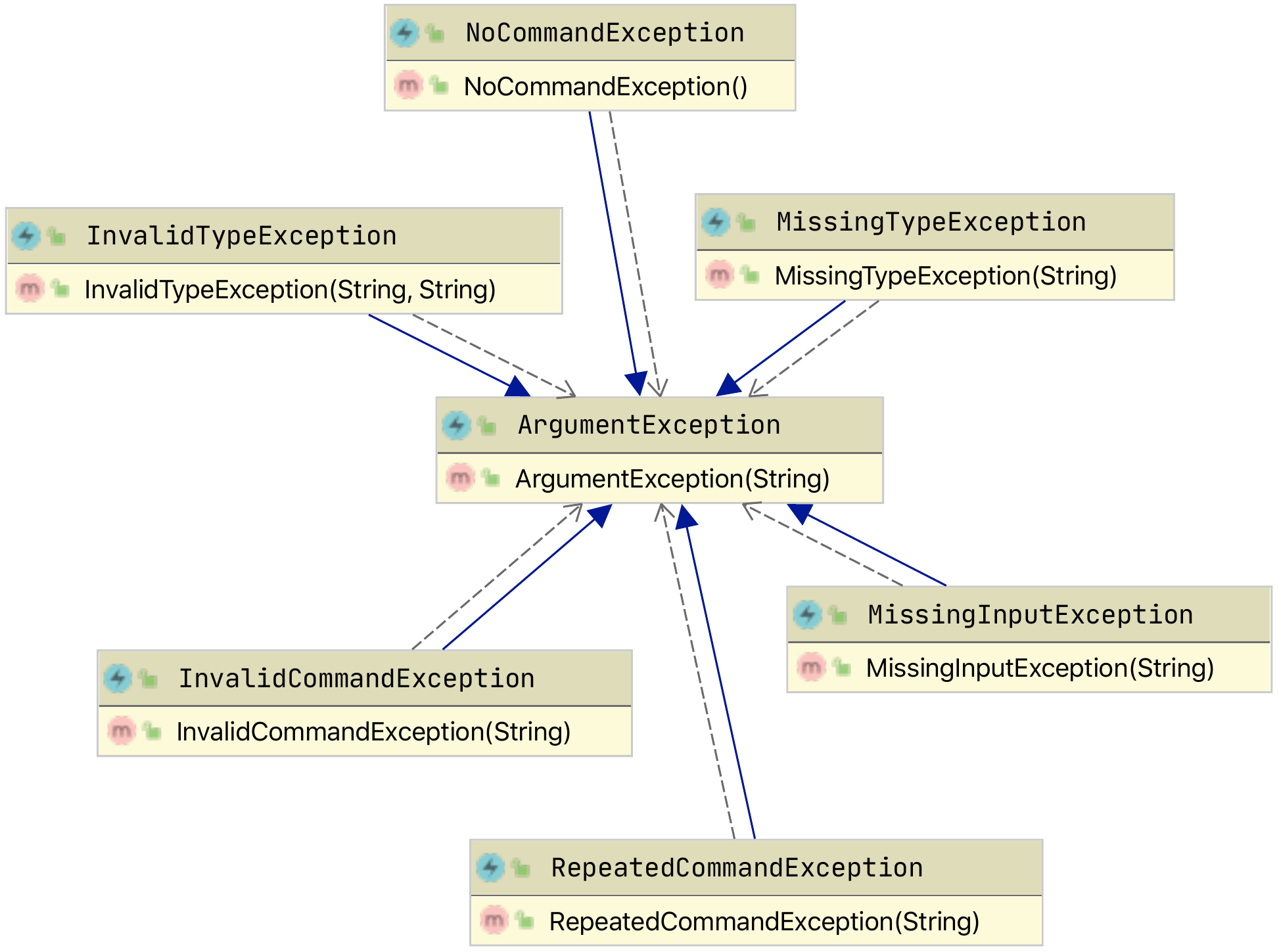}
    \caption{UML class diagram of the exception package.}
    \label{uml:exception}
\end{figure}

A custom exception class was defined specifically for each user related issue. These classes are all present in the \verb|exception| package and extend the \verb|ArgumentException| custom class. Aside from the \verb|MissingInputException| class, all of the other exception classes, namely the \verb|InvalidCommandException|, \verb|InvalidTypeException|, \verb|MissingTypeException|, \verb|NoCommandException|, and \verb|RepeatedCommandException| classes, can only be thrown during the argument parsing phase. Meanwhile, the \verb|MissingInputException| can only be thrown at the start of the execution of each command, because only then can it be established if there is a value for that input, be it either from the arguments or the previously executed commands. The \ac{UML} class diagram that represents this package can be seen in Figure \ref{uml:exception}.

\subsection{Logging}

There are log messages spread throughout the whole workflow, namely in the exception handling, the argument parsing, the command execution, and the data parsing, and together they all provide some useful information to the user, regarding the state of the execution of the workflow.

These messages are all logged through the same place, which is the \verb|Log| class from the \verb|log| package of the project. The main purpose of this class is to setup the logs according to a given configuration, and provide different types of log messages for different purposes. To do so, it wraps the Java built-in logger by using the \verb|Level|, \verb|LogManager|, and \verb|Logger| classes from the \verb|java.util.logging| built-in Java package. The configuration of the logs is given by the values inside the \verb|logging.properties| file in the \verb|resources| of the project. Some of the configurations available here include the handler, level, and format of the messages.

\begin{figure}[!ht]
    \centering
    \includegraphics[scale=0.25]{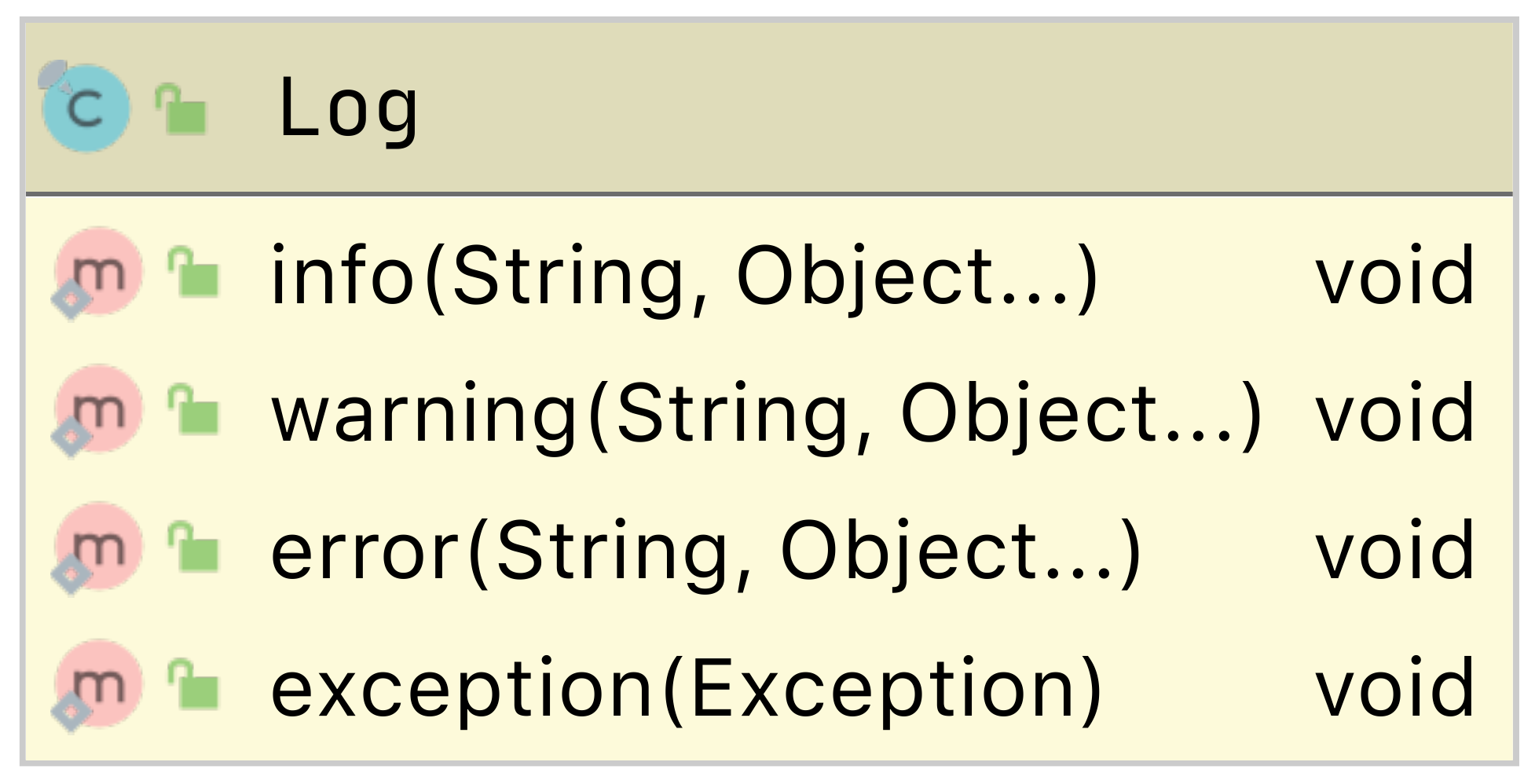}
    \caption{UML class diagram of the logging package.}
    \label{uml:logging}
\end{figure}

Each type of log message was translated into a different method inside the \verb|Log| class, namely the \verb|info|, \verb|warning|, \verb|error|, and \verb|exception| methods. The \verb|info| method is used to provide some workflow progression information, that is useful for the user to understand the state of the current execution. The \verb|warning| method is used for potential user input mistakes, that are not crucial to the progression of the workflow, since they can be ignored, such as duplicated, invalid or unused parameters. The \verb|error| method is only used in the \verb|Main| class, and is intended for logging only user input related issues that crucial to the progression of the workflow. Finally, the \verb|exception| method is only used in the \verb|Main| class as well, however its sole purpose is to log internal issues that need to be looked at by the developer, thus providing a stack trace of the thrown exception. The \ac{UML} class diagram that represents this package can be seen in Figure \ref{uml:logging}.

\section{Data}

The phylogenetic analysis workflow deals with three main types of data, namely datasets, distance matrices, and phylogenetic trees. Each step of the workflow always returns data of one data type, but may receive data of one or more data types. For that reason, instead of defining a different interface for each step, a common context object is used by the \verb|Main| class to aggregate all of the data and share it between all steps. That object is defined as the \verb|Context| class in the \verb|data| package, and it is responsible for storing the current values of each data type during each execution of the workflow.

\begin{figure}[!ht]
    \centering
    \includegraphics[scale=0.85]{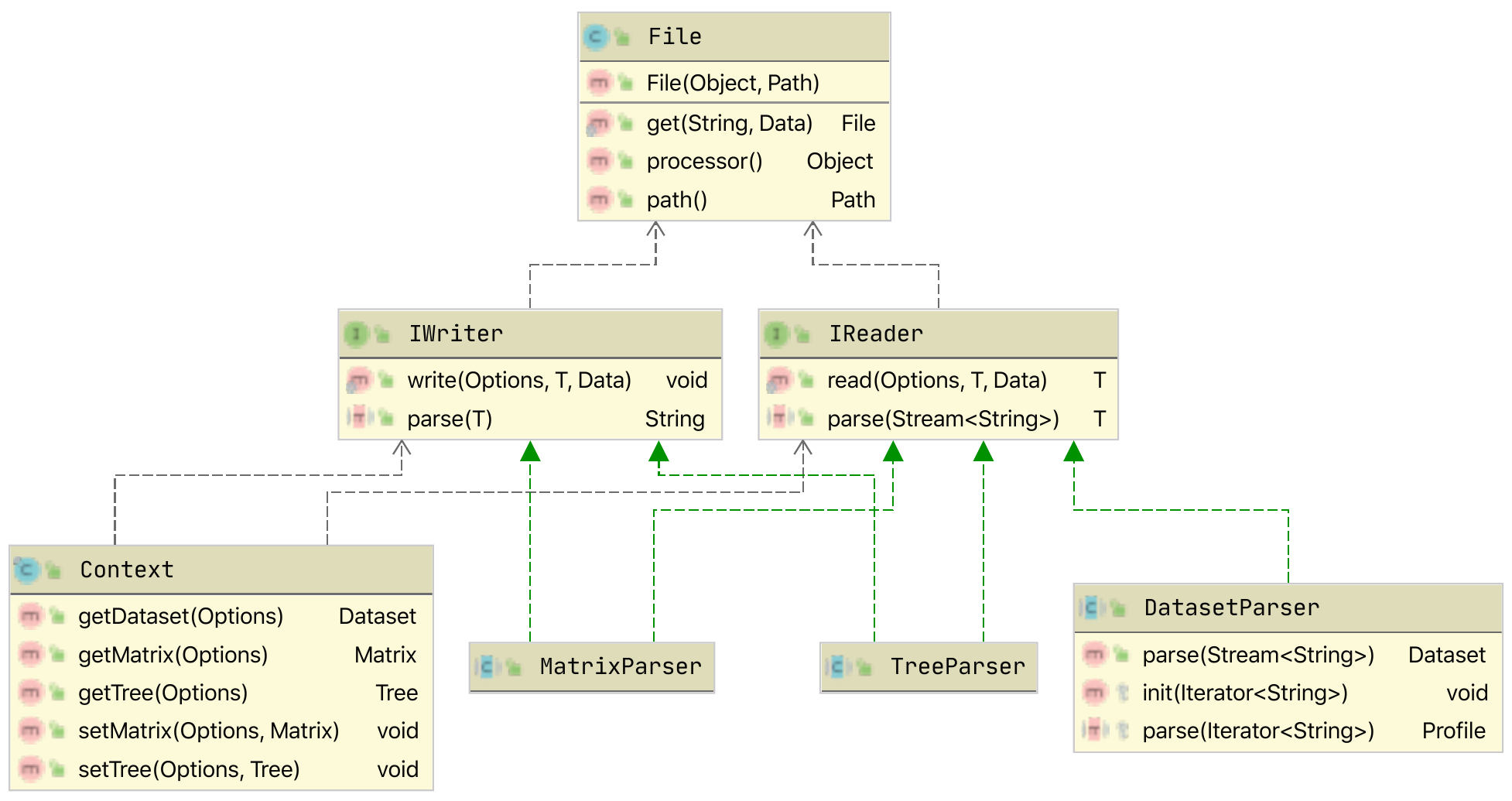}
    \caption{UML class diagram of the data package.}
    \label{uml:data}
\end{figure}

Each of the data types may be read from or written to a file, except for the dataset that can only be read from a file, as it is never an output of any step. This means that there are specific reading and writing concerns for certain data types in each step of the workflow. To ensure reusability and avoid repeated code, the reading and writing portions of the project are separated from the data parsing. The former are enclosed in two specific interfaces inside the \verb|data| package, namely the \verb|IReader| and \verb|IWriter| interfaces. Whereas the latter are enclosed in specific packages and classes for each data type, and include the parsing of files into data and vice versa. These are the \verb|DatasetParser|, \verb|MatrixParser|, and \verb|TreeParser| classes from the \verb|data.dataset|, \verb|data.matrix|, and \verb|data.tree| packages respectively, which implement the respective parsing methods of those interfaces. The \ac{UML} class diagram that represents this package can be seen in Figure \ref{uml:data}.

The \verb|File| class of this package is responsible for mapping the input and output parameters into data type implementations and files. If the user, for example, executes a command with the \verb|--dataset| option equal to \verb|snp:dataset.txt|, then this class, with the help of the \verb|Data| enum from the \verb|cli| package, will translate it into the reading of the file \verb|dataset.txt| with the \verb|snp| implementation of the \verb|DatasetParser| class.

\subsection{Dataset}

The initial data type used in the workflow is the dataset, which is composed of allelic profiles that define a species or taxa. Each of these profiles is represented by an identifier and several loci. Both of these data structures are represented internally in the \verb|data.dataset| package by the \verb|Dataset| and \verb|Profile| classes respectively.

\begin{figure}[!ht]
    \centering
    \includegraphics[scale=0.8]{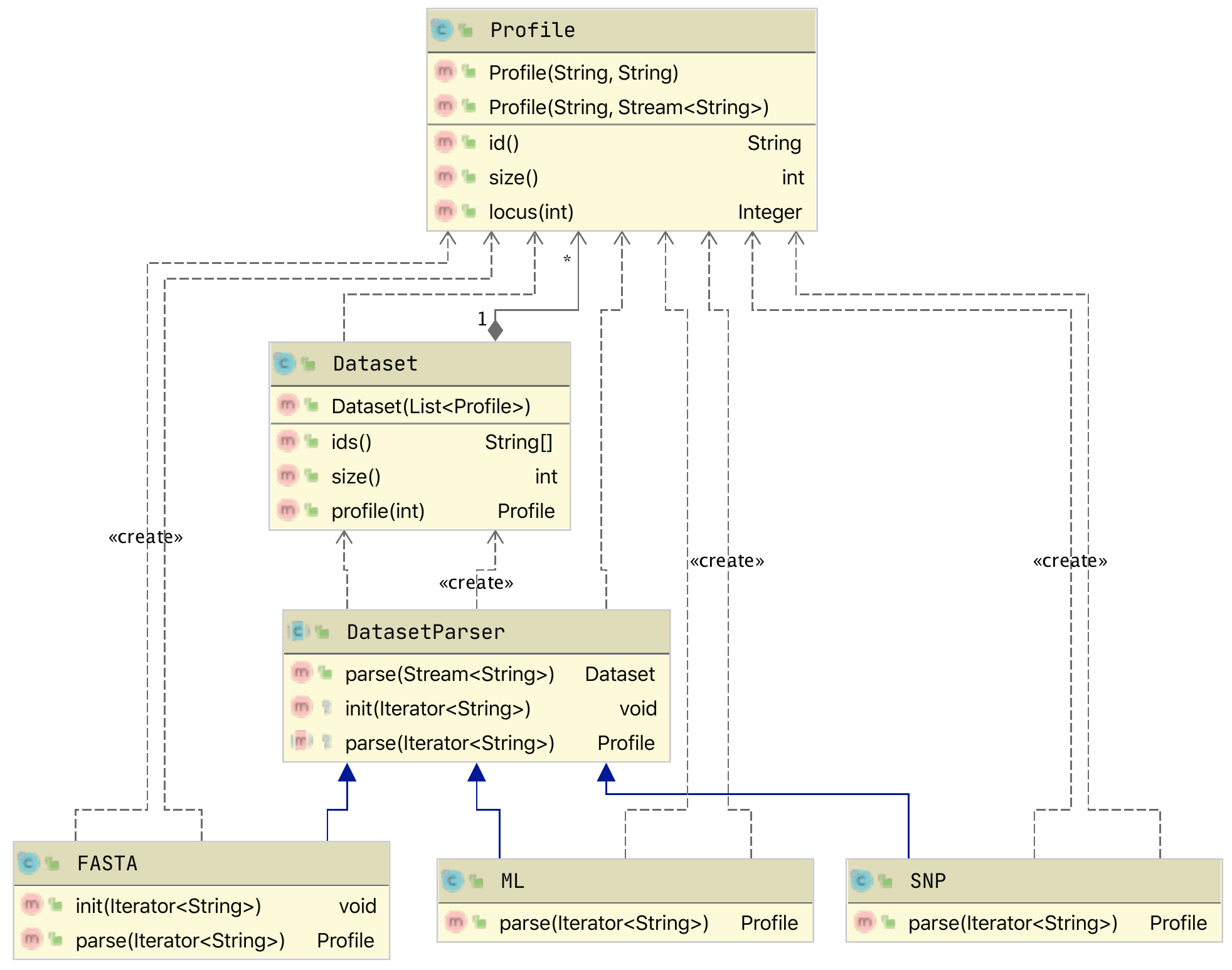}
    \caption{UML class diagram of the dataset package.}
    \label{uml:dataset}
\end{figure}

As previously mentioned, the dataset data type is only used in the phylogenetic analysis workflow as an input, and therefore its respective parser, which is represented by the \verb|DatasetParser| class in the \verb|data.dataset| package, only needs to implement the \verb|IReader| interface. This class is an abstract implementation of said interface, as it does not implement the whole dataset parsing concern, but simply puts in evidence some of the logic involved in parsing datasets in any format to ensure reusability and avoid repeated code. Therefore, each implementation of the dataset data type, namely the \verb|FASTA|, \verb|ML|, and \verb|SNP| classes, extend the \verb|DatasetParser| class. The \ac{UML} class diagram that represents this package can be seen in Figure \ref{uml:dataset}.

\subsection{Distance Matrix}

After the dataset data type, comes the distance matrix data type, which is composed of distances between each loci of a species or taxa that define the evolutionary distance between them. This data structure is represented in the \verb|data.matrix| package by the \verb|Matrix| class, which defines two inner functional interfaces for the distance calculation and correction, namely the \verb|IDistance| and \verb|ICorrection| interfaces, to be received as parameters to calculate and correct the distances of the matrix. This data type distinguishes between symmetric and asymmetric distance matrices to optimize its memory usage by only allocating the space required to store the different distances. Likewise, it only allocates space for the distances when it is necessary, relying on a lazy approach for the distance calculation.

\begin{figure}[!ht]
    \centering
    \includegraphics[scale=0.42]{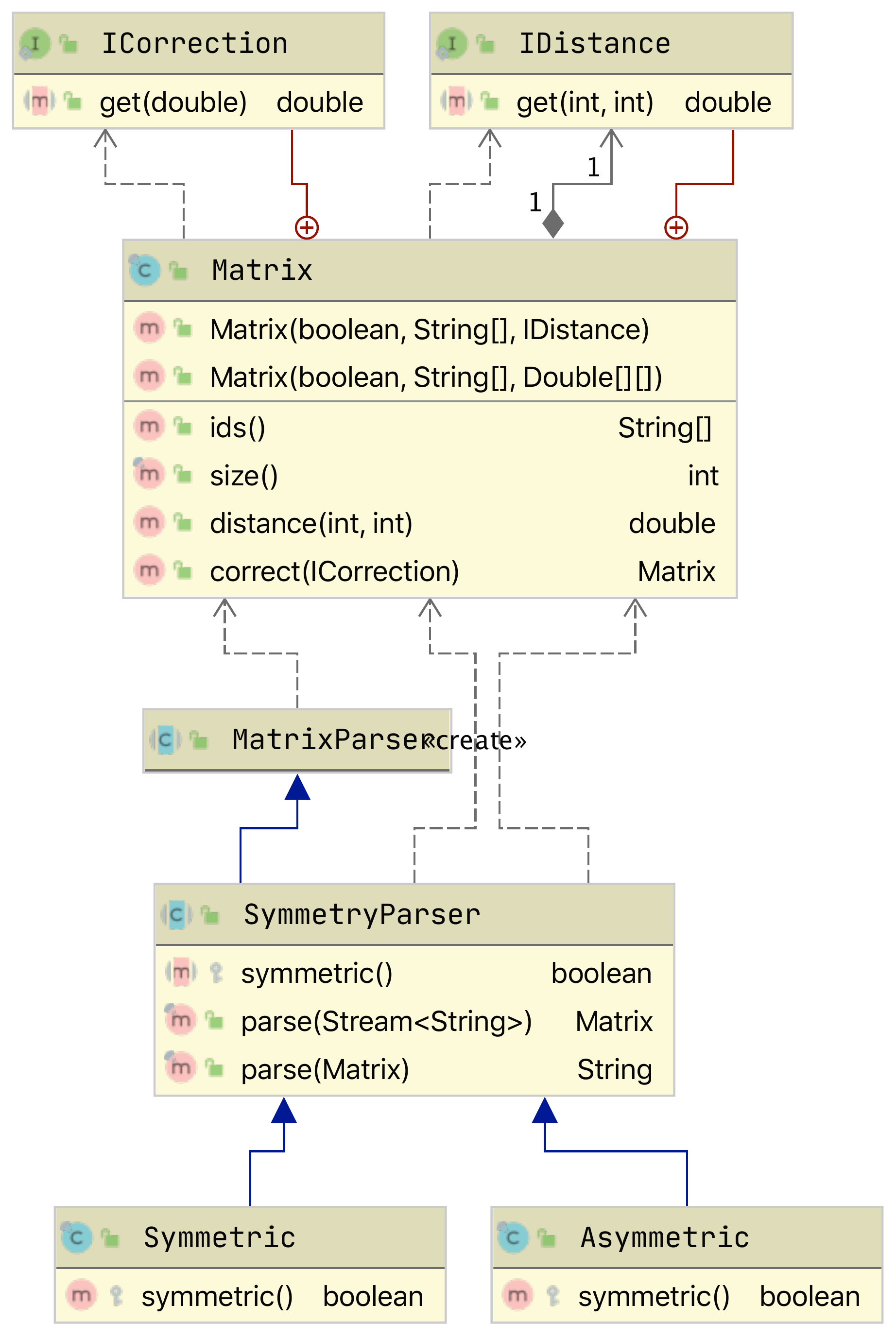}
    \caption{UML class diagram of the matrix package.}
    \label{uml:matrix}
\end{figure}

The matrix data type is used in the phylogenetic analysis workflow both as an input and an output, and therefore its respective parser, which is represented by the \verb|MatrixParser| class in the \verb|data.matrix| package, needs to implement both the \verb|IReader| and \verb|IWriter| interfaces. This class is an abstract implementation of said interface, as it does not implement any of the matrix parsing concern, but simply puts in evidence the fact that each matrix parser should implement both the \verb|IReader| and \verb|IWriter| interfaces. Therefore, each implementation of the data type, namely the \verb|Symmetric| and \verb|Asymmetric| classes, extend the \verb|MatrixParser| class. However, since these two formats are very similar, to ensure reusability and avoid repeated code, there is an intermediate abstract class that both extend, namely the \verb|SymmetryParser| class, which in turn extends the \verb|MatrixParser| class. The \ac{UML} class diagram that represents this package can be seen in Figure \ref{uml:matrix}.

\subsection{Phylogenetic Tree}

Lastly, after the distance matrix data type, comes the phylogenetic tree data type, which is composed of edges that represent the evolutionary relationships between the loci of the species or taxa. Each of these edges is represented by the identifiers of two loci and the distance between them. Both of these data structures are represented in the \verb|data.tree| package by the \verb|Tree| and \verb|Edge| classes respectively.

\begin{figure}[!ht]
    \centering
    \includegraphics[scale=0.5]{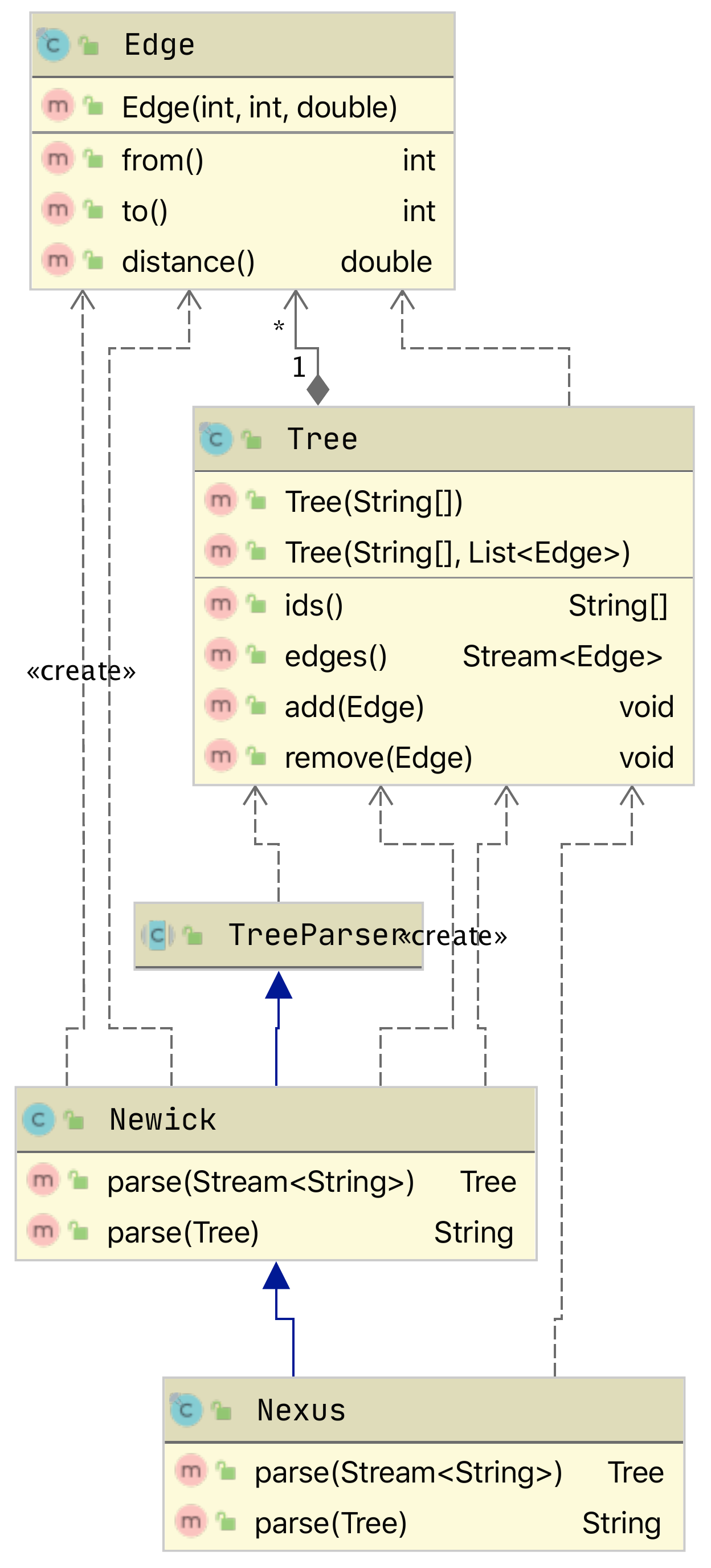}
    \caption{UML class diagram of the tree package.}
    \label{uml:tree}
\end{figure}

The tree data type is used in the phylogenetic analysis workflow both as an input and an output, and therefore its respective parser, which is represented by the \verb|TreeParser| class in the \verb|data.tree| package, needs to implement both the \verb|IReader| and \verb|IWriter| interfaces. This class is an abstract implementation of said interface, as it does not implement any of the tree parsing concern, but simply puts in evidence the fact that each tree parser should implement both the \verb|IReader| and \verb|IWriter| interfaces. Therefore, each implementation of the tree data type, namely the \verb|Newick| and \verb|Nexus| classes, extend the \verb|TreeParser| class. However, since the Nexus format is very similar to the Newick format, in the sense that it only adds information on top of it, to ensure reusability and avoid repeated code, the \verb|Nexus| class extends the \verb|Newick| class. The \ac{UML} class diagram that represents this package can be seen in Figure \ref{uml:tree}.

\section{Commands}

The phylogenetic analysis workflow can be decomposed into four consecutive steps, the distance calculation, distance correction, inference algorithm, and local optimization steps. The \verb|Main| class setups the workflow by calling the \verb|ICommand| interface from the \verb|command| package for each step of the workflow. For each call to this interface it takes the arguments and context it receives to instantiate the implementations of the corresponding command and input and output data parsers, and then gets the input for that command, executes it, and stores its result. If the command is repeatable, such as the local optimization, it will repeat this logic until there are no more \ac{CLI} arguments for that command. However, if the command is optional, such as the distance correction and local optimization, and is not provided in the \ac{CLI} arguments it will be skipped. Despite the workflow being correctly setup by the \verb|Main| class, it may not be executed in its entirety and some steps may be skipped as the user may choose to ignore some steps of the workflow by providing a file as input to the steps that needed the output of other steps.

\begin{figure}[!ht]
    \centering
    \includegraphics[scale=0.8]{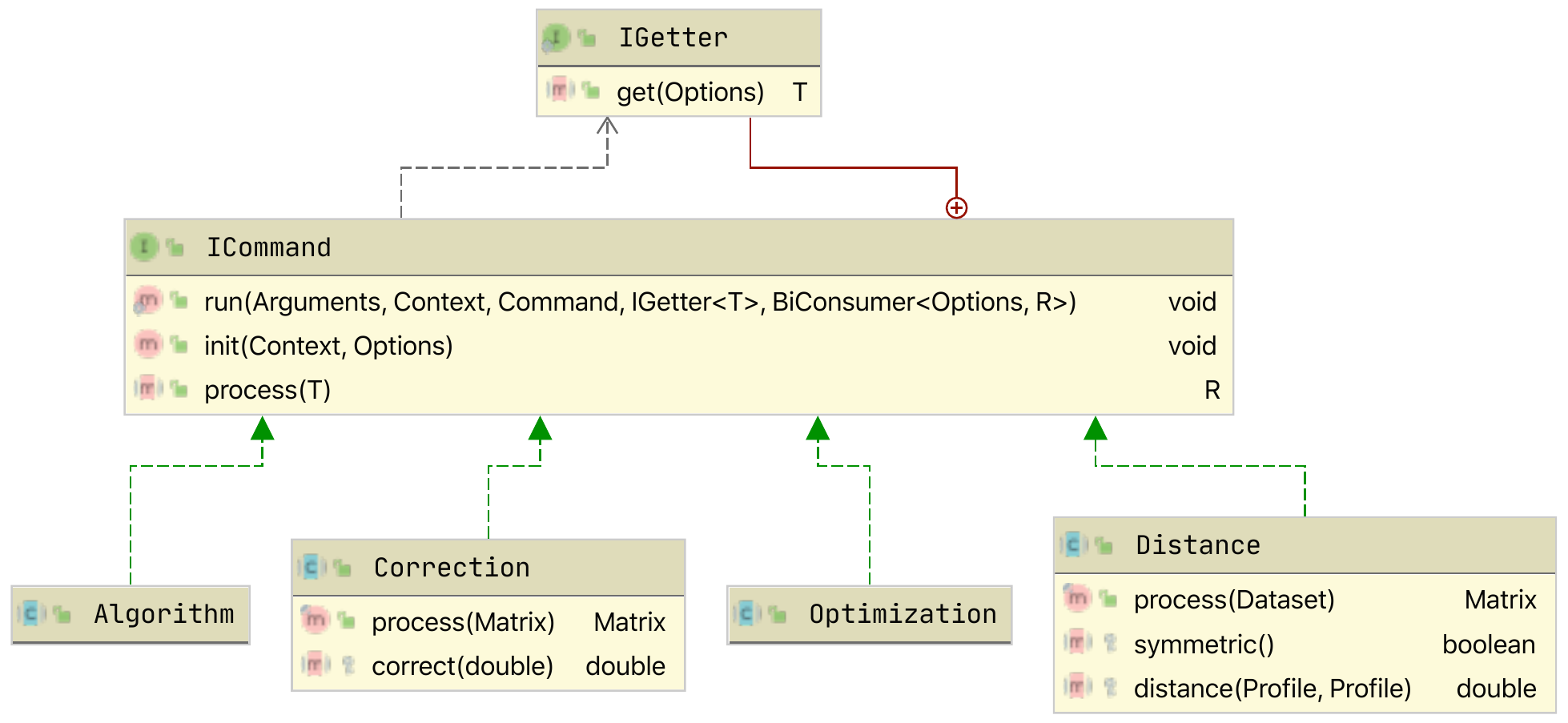}
    \caption{UML class diagram of the command package.}
    \label{uml:command}
\end{figure}

For each of the four steps of the workflow there is an implementation of the \verb|ICommand| interface, respectively the \verb|Distance|, \verb|Correction|, \verb|Algorithm|, and \verb|Optimization| abstract classes from the \verb|command.distance|, \verb|command.correction|, \verb|command.algorithm|, and \verb|command.optimization| packages respectively. The implementations of these abstract classes are what the \verb|ICommand| interface instantiates and runs. Each implementation of these abstract classes must directly or indirectly implement the only abstract method the interface provides, which is responsible for the logic that transforms the input data into the output data. By default, this interface only provides one input to the execution, however it provides another method, which is not abstract, that each implementation may override to parse additional options it might need for its execution. The \ac{UML} class diagram that represents this package can be seen in Figure \ref{uml:command}.

The \verb|IGetter| interface, internal to the \verb|ICommand| interface, is a functional interface similar to the Java built-in \verb|Function| interface, with the difference that it may throw an exception related to a missing input. It is used to represent the retrieval of the input data from the arguments or context, and it may throw an exception if it can not retrieve a value from either. Whereas the Java built-in \verb|BiConsumer| class is used to represent the storage of the output in the context and in an output file if any, without throwing any exception.

\subsection{Distance Calculation}

\begin{figure}[!ht]
    \centering
    \includegraphics[scale=0.85]{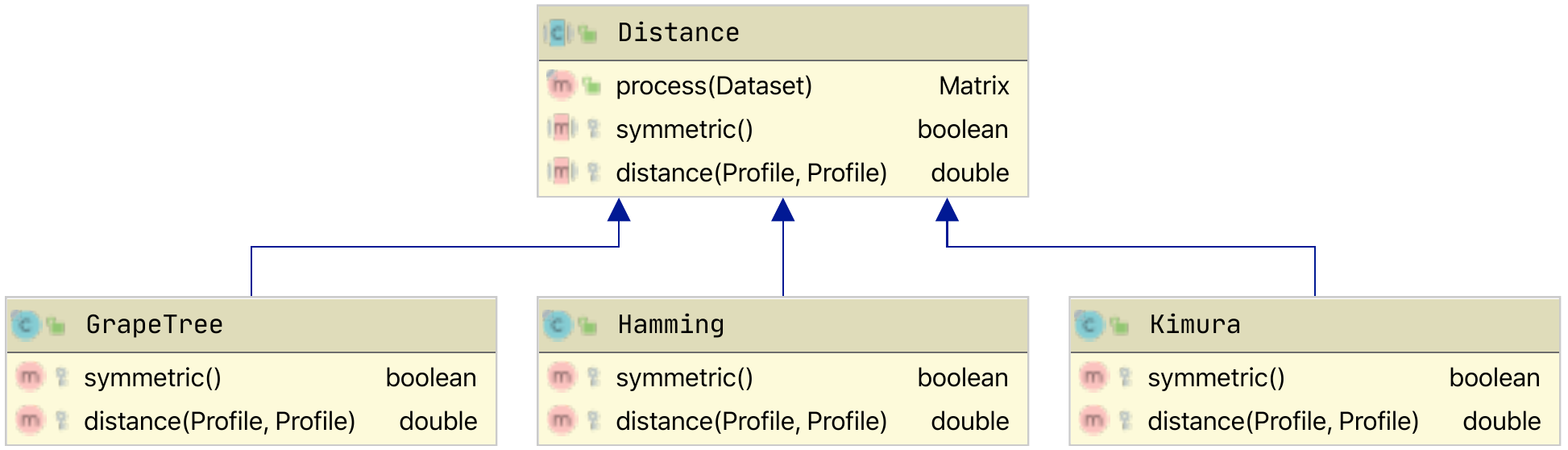}
    \caption{UML class diagram of the distance package.}
    \label{uml:distance}
\end{figure}

The logic of the distance calculation step of the phylogenetic analysis workflow is provided by the \verb|Distance| class from the \verb|command.distance| package. This class is an abstract implementation of the \verb|ICommand| interface that receives a dataset and transforms it into a distance matrix with the evolutionary distances between the profiles calculated according to a distance metric. Each implementation of this abstract class only has to define whether the resulting distance matrix is symmetric or not, and provide a method to calculate the evolutionary distance between any two given profiles. The available implementations of this abstract class are the \verb|GrapeTree|, \verb|Hamming|, and \verb|Kimura| classes. The \ac{UML} class diagram that represents this package can be seen in Figure \ref{uml:distance}.

\subsection{Distance Correction}

\begin{figure}[!ht]
    \centering
    \includegraphics[scale=0.21]{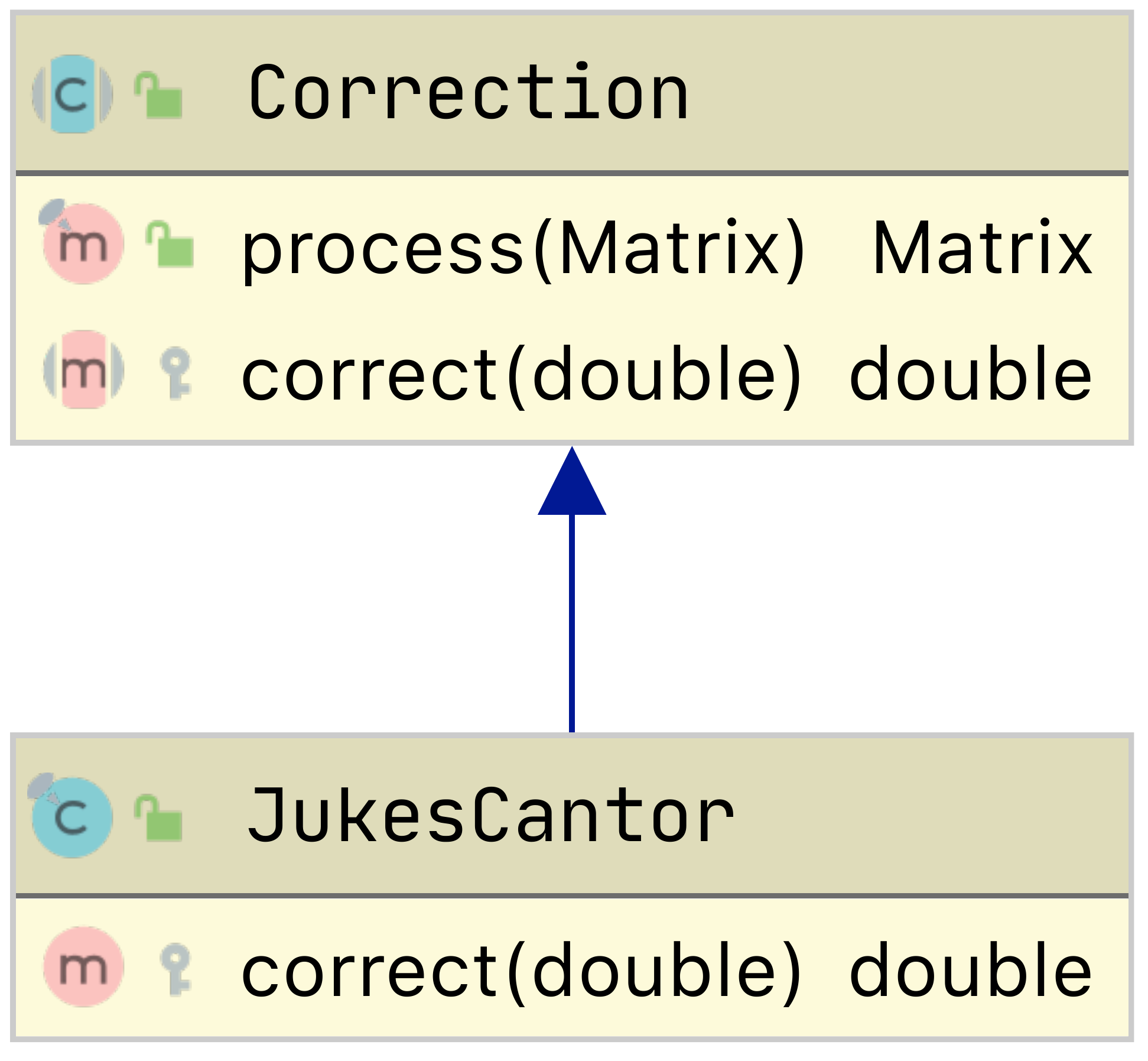}
    \caption{UML class diagram of the correction package.}
    \label{uml:correction}
\end{figure}

After the distance calculation step comes the optional distance correction step of the phylogenetic analysis workflow. The logic of this step is provided by the \verb|Correction| class from the \verb|command.correction| package. This class is an abstract implementation of the \verb|ICommand| interface that receives a distance matrix and transforms it into another distance matrix with the evolutionary distances corrected according to a model of evolution. Each implementation of this abstract class only has to provide a method to correct any given evolutionary distance. The only available implementation of this abstract class is the \verb|JukesCantor| class. The \ac{UML} class diagram that represents this package can be seen in Figure \ref{uml:correction}.

\subsection{Inference Algorithm}

\begin{figure}[!ht]
    \centering
    \includegraphics[scale=0.85]{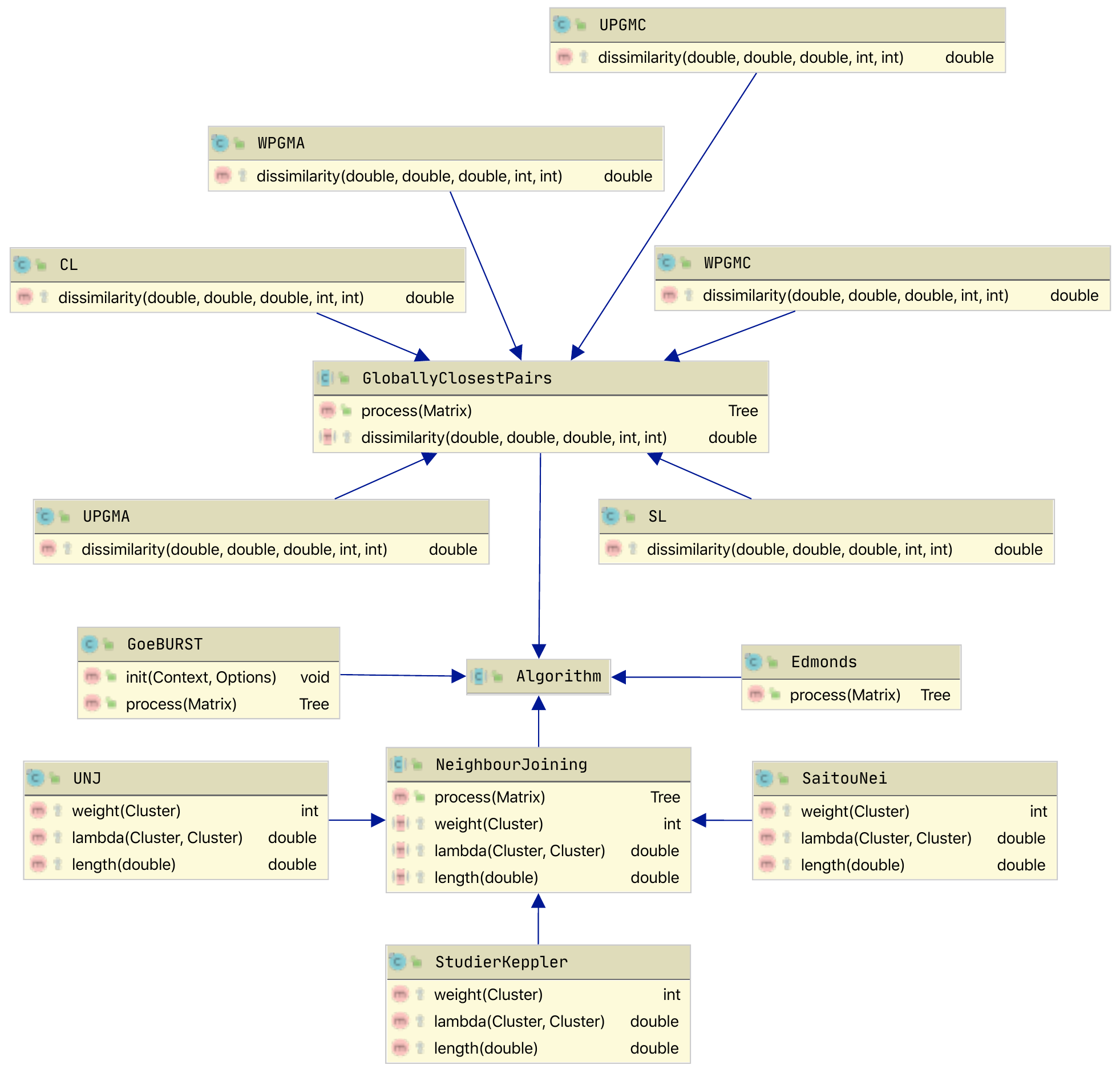}
    \caption{UML class diagram of the algorithm package.}
    \label{uml:algorithm}
\end{figure}

The logic of the inference algorithm step of the phylogenetic analysis workflow is provided by the \verb|Algorithm| class from the \verb|command.algorithm| package. This class is an abstract implementation of the \verb|ICommand| interface that receives a distance matrix and transforms it into a phylogenetic tree with the evolutionary relationships between the profiles selected by a clustering algorithm. Despite sharing the same general scheme, each clustering algorithm can be better optimized by using its own scheme and data structures. Thus to achieve a better performance, each implementation of this abstract class must define how the distance matrix is transformed into a phylogenetic tree. The \ac{UML} class diagram that represents this package can be seen in Figure \ref{uml:algorithm}.

The \verb|Algorithm| abstract class is directly implemented by the \verb|GloballyClosestPairs|, \verb|GoeBURST|, \verb|Edmonds|, and \verb|NeighbourJoining| classes from the respective \verb|command.algorithm.gcp|, \verb|command.algorithm.goeburst|, \verb|command.algorithm.edmonds|, and \verb|command.algorithm.nj| packages. The \verb|GloballyClosestPairs| class is an abstract implementation of the \verb|Algorithm| class, which concentrates the overall logic of \ac{GCP} algorithms in just one place. The available implementations of this abstract class are the \verb|SL|, \verb|CL|, \verb|UPGMA|, \verb|WPGMA|, \verb|UPGMC|, and \verb|WPGMC| classes, which only have to implement one method to obtain the dissimilarity between two nodes of the phylogenetic tree, as it is their only difference. The \verb|GoeBURST| and \verb|Edmonds| classes are final implementations of the \verb|Algorithm| class. Despite both being \ac{MST} algorithms, their implementations do not share a common logic between them as they can be better optimized if implemented completely separately. Finally, the \verb|NeighbourJoining| class is an abstract implementation of the \verb|Algorithm| class, which concentrates the overall logic of \ac{NJ} algorithms in just one place. The available implementations of this abstract class are the \verb|SaitouNei| and \verb|StudierKeppler| classes, which have to implement three methods to obtain the weight of a cluster, the proportion of a given cluster to another, and the length corresponding to a given distance, as it is where they differ.

\subsection{Local Optimization}

\begin{figure}[!ht]
    \centering
    \includegraphics[scale=0.35]{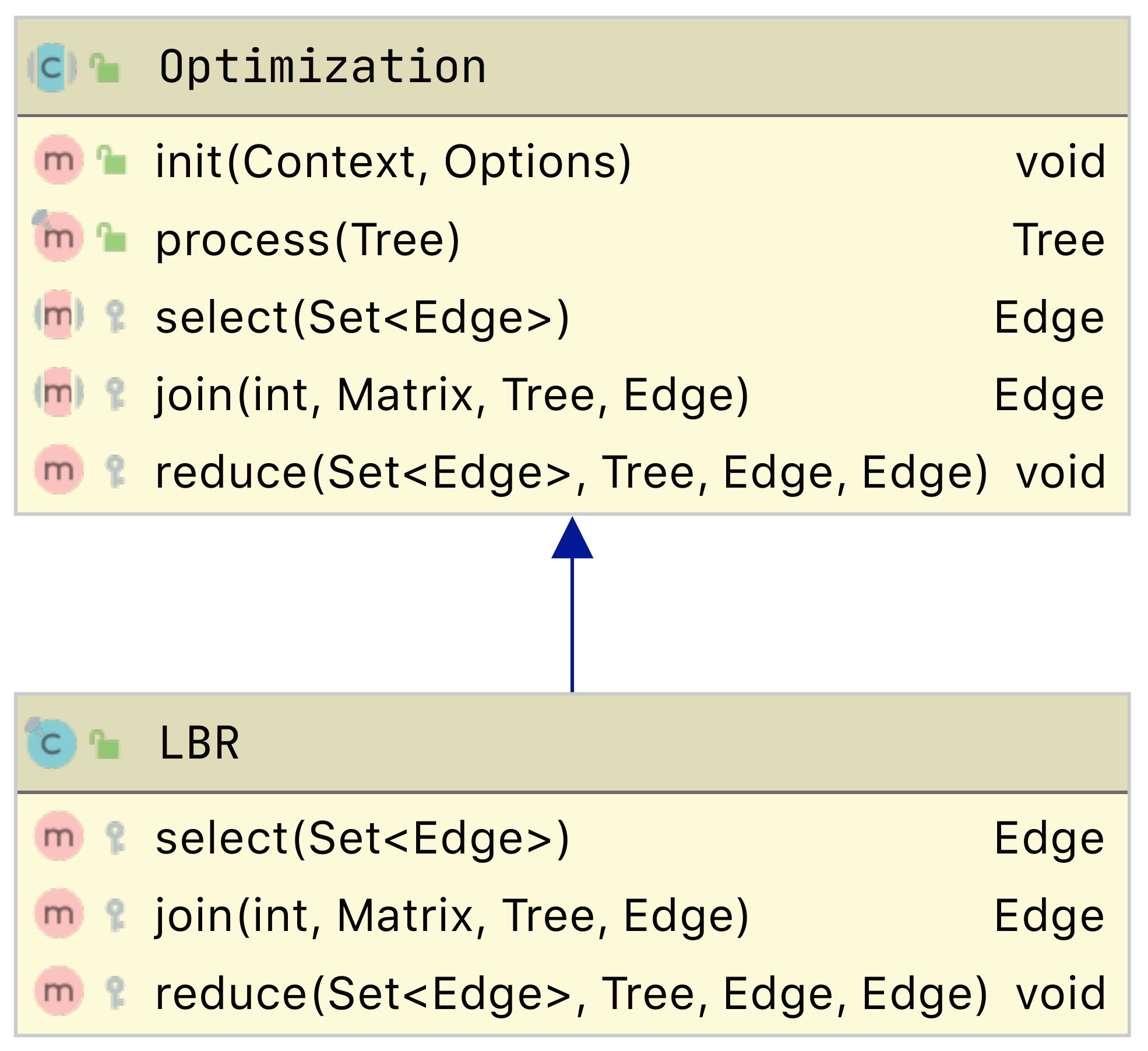}
    \caption{UML class diagram of the optimization package.}
    \label{uml:optimization}
\end{figure}

After the inference algorithm step comes the optional local optimization step of the phylogenetic analysis workflow. The logic of this step is provided by the \verb|Optimization| class from the \verb|command.optimization| package. This class is an abstract implementation of the \verb|ICommand| interface that receives a phylogenetic tree and transforms it into another phylogenetic tree with the evolutionary relationships locally optimized according to an algorithm. The general scheme of local optimization algorithms is provided by this abstract class, and therefore each of its implementations only has to implement the selection and joining steps of the algorithm which is where they differ. The only available implementation of this abstract class is the \verb|LBR| class. This class however also overrides the reduction step as it is a particularity of the \ac{LBR} algorithm. The \ac{UML} class diagram that represents this package can be seen in Figure \ref{uml:optimization}.

\section{Discussion}

This project is composed of six main concerns, namely the \ac{CLI} argument parsing, mapping of arguments to commands and data parsers, exception handling, logging, data parsing, and command execution. Each of these concerns is translated into a different package inside the \verb|pt.ist.phylolib| package, respectively the \verb|cli|, \verb|reflection|, \verb|exception|, \verb|logging|, \verb|data|, and \verb|command| packages.

The implementation of this project takes into account reusability, and for that reason the commands and data types concentrate as much reusable code as possible in common hierarchical classes. As a result, it also improves its extensibility, as it becomes easier to extend commands and data types since most of the code necessary to implement a command or data type is already written. Nonetheless, performance is also taken into account in the implementation, as some code that could be reused is not, due to optimizations that can be made specifically to some algorithm implementations.

\chapter{Experimental Evaluation}

The purpose of this chapter is to enumerate and explain all of the tests that were performed on this library and the results that were obtained from their executions, as well as the environment in which they were performed and the constants that were used.

All components of this library were tested in terms of their functionality, through unit testing implemented using the TestNG \cite{testng} framework, except for the local optimization which does not have another implementation to compare its results to. However, only the algorithm component was tested in terms of time and memory performances, since it is the core operation of the workflow and the one that requires the most time and memory to execute. This chapter will focus itself on the time and memory performances, comparing both the performance of each algorithm against each other, as well as the performance of each algorithm with an implicit matrix versus with an explicit matrix.

Using an implicit matrix in the execution of an algorithm is another way of saying to only calculate the distances from the dataset as the algorithm requests them, instead of already having them precomputed in a distance matrix. This is also known as the lazy version. While using an explicit matrix is the opposite of that, that is, to already have the distances precomputed in a distance matrix. This is also known as the eager version. The lazy and eager versions can be translated into executing the library providing a dataset and a distance matrix respectively. Comparing the results of the lazy and eager versions is useful to help better understand the advantages and disadvantages of storing the distance matrix and reusing it.

The time and memory performances were tested through the implementation and execution of benchmarks with 10 warmups and 20 iterations, over the first ten to one thousand profiles of the Streptococcus pneumoniae dataset \cite{pubmlst}, using the Hamming distance as the distance calculation method. The results are represented as a function of the number of profiles $n$. The same dataset and distance calculation method were used throughout the benchmarks to provide an equal and fair evaluation to all algorithms. For that same reason, all tests were performed in the same machine, in this case with a 2.6 GHz 6-Core Intel Core i7 processor and a 16 GB 2667 MHz DDR4 memory.

\section{Time}

This section analyzes the results obtained from the time performance benchmarks, comparing the time complexity of each algorithm, as well as the difference in time performance of the eager and lazy versions.

The average running time that each algorithm took to execute in the eager version, over the increasing number of profiles of the given dataset, is represented in Table \ref{table:eagertime}, in milliseconds. From this table it is possible to see the difference in time complexity that exists between the \ac{NJ} algorithms and all the others.

\begin{table}[!ht]
    \centering
\begin{tabular}{ | c | c c c c c c c c c c c | }
        \hline
        \diagbox{Algorithm}{Profiles} & 10 & 100 & 200 & 300 & 400 & 500 & 600 & 700 & 800 & 900 & 1000 \\
        \hline
        goeBURST & 0 & 1 & 2 & 4 & 6 & 10 & 14 & 21 & 31 & 40 & 53 \\

        goeBURST Full & 0 & 5 & 24 & 64 & 124 & 213 & 321 & 442 & 616 & 794 & 1005 \\
        
        Edmonds & 0 & 15 & 79 & 206 & 381 & 666 & 971 & 1361 & 1844 & 2371 & 2996 \\
        
        CL & 0 & 4 & 22 & 59 & 110 & 193 & 284 & 399 & 500 & 658 & 772 \\
        
        SL & 0 & 2 & 13 & 35 & 65 & 111 & 163 & 235 & 301 & 383 & 468 \\
        
        UPGMA & 0 & 4 & 22 & 59 & 115 & 200 & 296 & 424 & 544 & 705 & 828 \\
        
        UPGMC & 0 & 1 & 7 & 31 & 55 & 84 & 138 & 199 & 267 & 336 & 408 \\
        
        WPGMA & 0 & 4 & 22 & 61 & 119 & 202 & 288 & 425 & 545 & 714 & 852 \\
        
        WPGMC & 0 & 2 & 12 & 34 & 68 & 119 & 173 & 250 & 327 & 386 & 510 \\
        
        Saitou Nei & 0 & 9 & 82 & 232 & 574 & 1125 & 2051 & 3362 & 5738 & 8261 & 11538 \\
        
        Studier Keppler & 0 & 10 & 86 & 232 & 555 & 1118 & 2058 & 3379 & 5732 & 8272 & 11593 \\
        
        UNJ & 0 & 9 & 87 & 230 & 559 & 1082 & 1976 & 3257 & 5691 & 7822 & 10828 \\
        \hline
    \end{tabular}
    \caption{Average running times in milliseconds for ten to one thousand profiles using the eager version.}
    \label{table:eagertime}
\end{table}

\begin{figure}[!ht]
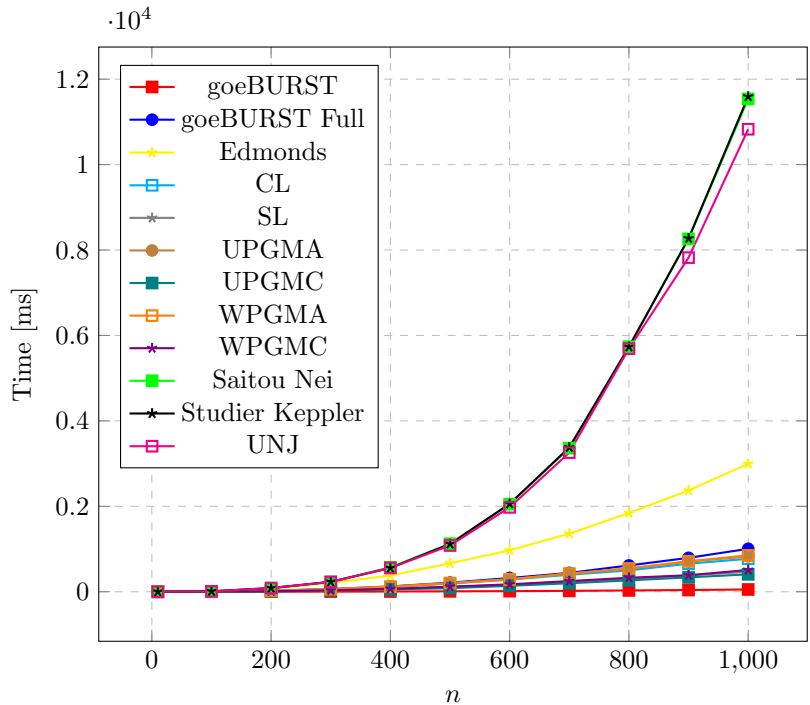

    \centering
    %
    \caption{Running times in milliseconds for ten to one thousand profiles using the eager version.}
    \label{graph:time}
\end{figure}

The obtained results can also be represented in the form of a plot graph, as in Figure \ref{graph:time}, where it is possible to better understand the differences in time complexity between the different types of algorithms.

Despite having the same time complexity as other \ac{MST} and \ac{GCP} algorithms, it is possible to see from the previous plot graph that the Edmonds algorithm is much slower than the others.

The average running time that each algorithm took to execute in the lazy version, over the increasing number of profiles of the given dataset, is represented in Table \ref{table:lazytime}, in milliseconds.

\begin{table}[!ht]
    \centering
    \begin{tabular}{ | c | c c c c c c c c c c c | }
        \hline
        \diagbox{Algorithm}{Profiles} & 10 & 100 & 200 & 300 & 400 & 500 & 600 & 700 & 800 & 900 & 1000 \\
        \hline
        goeBURST & 0 & 1 & 2 & 5 & 13 & 22 & 30 & 40 & 53 & 75 & 92 \\

        goeBURST Full & 0 & 5 & 24 & 68 & 130 & 233 & 342 & 473 & 641 & 819 & 1044 \\
        
        Edmonds & 0 & 17 & 82 & 209 & 403 & 674 & 991 & 1365 & 1876 & 2384 & 3077 \\
        
        CL & 0 & 4 & 24 & 62 & 121 & 193 & 300 & 378 & 529 & 663 & 811 \\
        
        SL & 0 & 2 & 16 & 39 & 65 & 101 & 178 & 244 & 301 & 396 & 463 \\
        
        UPGMA & 0 & 4 & 24 & 62 & 128 & 203 & 311 & 416 & 576 & 718 & 881 \\
        
        UPGMC & 0 & 2 & 13 & 35 & 58 & 88 & 149 & 221 & 278 & 350 & 413 \\
        
        WPGMA & 0 & 4 & 25 & 64 & 123 & 208 & 326 & 427 & 578 & 728 & 883 \\
        
        WPGMC & 0 & 2 & 15 & 37 & 71 & 105 & 189 & 262 & 343 & 401 & 506 \\
        
        Saitou Nei & 0 & 9 & 89 & 227 & 574 & 1178 & 2001 & 3236 & 5442 & 7827 & 10981 \\
        
        Studier Keppler & 0 & 10 & 89 & 228 & 575 & 1107 & 1995 & 3235 & 5435 & 7868 & 10990 \\
        
        UNJ & 0 & 9 & 87 & 226 & 565 & 1102 & 1944 & 3137 & 5190 & 7449 & 10341 \\
        \hline
    \end{tabular}
    \caption{Running times in milliseconds for ten to one thousand profiles using the lazy version.}
    \label{table:lazytime}
\end{table}

The difference in the results of the eager and lazy versions can be better compared in individual plot graphs. Figures \ref{graph:msttime}, \ref{graph:gcptime} and \ref{graph:njtime} represent these differences for \ac{MST}, \ac{GCP} and \ac{NJ} algorithms respectively.

\begin{figure}[!ht]
    \centering
    \begin{subfigure}[b]{0.49\textwidth}
        %
    \end{subfigure}
    \caption{Running times in milliseconds for MST algorithms compared to their time complexity.}
    \label{graph:msttime}
\end{figure}

\begin{figure}[!ht]
    \vspace*{0.85cm}
    \centering
    \begin{subfigure}[b]{0.49\textwidth}
        %
    \end{subfigure}
    \caption{Running times in milliseconds for GCP algorithms compared to their time complexity.}
    \label{graph:gcptime}
\end{figure}

\begin{figure}[!ht]
    \centering
    \begin{subfigure}[b]{0.49\textwidth}
        %
    \end{subfigure}
    \caption{Running times in milliseconds for NJ algorithms compared to their time complexity.}
    \label{graph:njtime}
\end{figure}

From these individual plot graphs it is possible to see that the implementations of the algorithms conform to their theoretical time complexities, namely $\mathcal{O}(n^3)$ for \ac{NJ} algorithms and $\mathcal{O}(n^2)$ for \ac{MST} and \ac{GCP} algorithms. It is also possible to see a difference in running time between the eager and lazy versions of the algorithms. However, these differences are hardly noticeable due to the small impact that the distance calculation has on the workflow, compared to the inference algorithm, except when compared to \ac{goeBURST}, as its running times are small enough to notice a difference. Thus, the benefit of storing the distance matrix in a file and reusing it is almost insignificant in terms of running time for most algorithms.

\section{Memory}

The implementation of the memory performance benchmarks relied on the \verb|MemoryPoolMXBean| interface from the \verb|java.lang.management| package, which represents a management interface of the memory resources managed by the \ac{JVM}. By using this interface it was possible to get the peak of memory usage of a memory pool since the virtual machine was started.

The average memory usage that each algorithm took to execute in the eager version, over the increasing number of profiles of the given dataset, is represented in Table \ref{table:eagermemory}, in megabytes. From this table it is possible to see that the \ac{goeBURST} algorithms require the lesser memory between all algorithms, while the Edmonds algorithm requires the most.

\begin{table}[!ht]
    \centering
\begin{tabular}{ | c | c c c c c c c c c c c | }
        \hline
        \diagbox{Algorithm}{Profiles} & 10 & 100 & 200 & 300 & 400 & 500 & 600 & 700 & 800 & 900 & 1000 \\
        \hline
        goeBURST & 11 & 11 & 13 & 16 & 19 & 23 & 29 & 35 & 42 & 51 & 60 \\

        goeBURST Full & 13 & 13 & 16 & 22 & 27 & 34 & 47 & 59 & 71 & 82 & 95 \\
        
        Edmonds & 12 & 19 & 34 & 73 & 103 & 131 & 174 & 216 & 285 & 400 & 435 \\
        
        CL & 13 & 14 & 20 & 33 & 44 & 71 & 88 & 126 & 159 & 220 & 219 \\
        
        SL & 13 & 14 & 20 & 28 & 39 & 55 & 75 & 93 & 117 & 139 & 187 \\
        
        UPGMA & 12 & 14 & 20 & 29 & 44 & 73 & 90 & 122 & 158 & 207 & 211 \\
        
        UPGMC & 13 & 14 & 19 & 26 & 38 & 55 & 70 & 100 & 122 & 146 & 156 \\
        
        WPGMA & 12 & 14 & 20 & 31 & 52 & 81 & 87 & 118 & 155 & 215 & 209 \\
        
        WPGMC & 13 & 14 & 19 & 25 & 38 & 52 & 73 & 89 & 113 & 136 & 166 \\
        
        Saitou Nei & 8 & 18 & 78 & 105 & 159 & 215 & 280 & 324 & 331 & 349 & 358 \\
        
        Studier Keppler & 13 & 21 & 92 & 112 & 160 & 211 & 282 & 335 & 336 & 354 & 373 \\
        
        UNJ & 10 & 19 & 90 & 106 & 161 & 218 & 287 & 353 & 356 & 384 & 389 \\
        \hline
    \end{tabular}
    \caption{Peak memory usage in megabytes for ten to one thousand profiles using the eager version.}
    \label{table:eagermemory}
\end{table}

\begin{figure}[!ht]
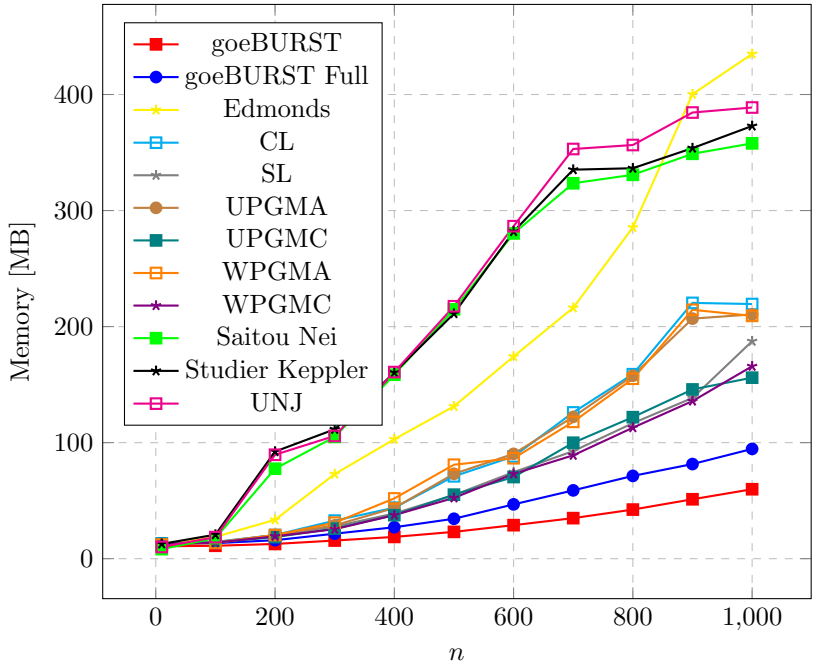

    \centering
    %
    \caption{Peak memory usage in megabytes for ten to one thousand profiles using the eager version.}
    \label{graph:memory}
\end{figure}

The obtained results can also be represented as a plot graph, as in Figure \ref{graph:memory}, where it is possible to better understand the differences in memory complexity between the different implemented algorithms.

The average memory usage that each algorithm took to execute in the lazy version, over the increasing number of profiles of the given dataset, is represented in Table \ref{table:lazymemory}, in megabytes.

\begin{table}[!ht]
    \centering
    \begin{tabular}{ | c | c c c c c c c c c c c | }
        \hline
        \diagbox{Algorithm}{Profiles} & 10 & 100 & 200 & 300 & 400 & 500 & 600 & 700 & 800 & 900 & 1000 \\
        \hline
        goeBURST & 11 & 11 & 13 & 16 & 18 & 20 & 24 & 30 & 39 & 51 & 58 \\

        goeBURST Full & 13 & 13 & 15 & 22 & 27 & 40 & 49 & 58 & 77 & 84 & 98 \\
        
        Edmonds & 12 & 18 & 36 & 71 & 102 & 131 & 171 & 222 & 296 & 370 & 438 \\
        
        CL & 13 & 14 & 21 & 35 & 54 & 82 & 100 & 124 & 161 & 196 & 246 \\
        
        SL & 13 & 14 & 21 & 29 & 45 & 66 & 79 & 99 & 137 & 162 & 189 \\
        
        UPGMA & 13 & 14 & 20 & 32 & 54 & 77 & 94 & 126 & 160 & 209 & 237 \\
        
        UPGMC & 13 & 14 & 20 & 26 & 39 & 56 & 74 & 99 & 130 & 146 & 173 \\
        
        WPGMA & 12 & 13 & 20 & 34 & 46 & 73 & 101 & 129 & 163 & 188 & 239 \\
        
        WPGMC & 13 & 14 & 20 & 26 & 41 & 59 & 76 & 99 & 129 & 142 & 173 \\
        
        Saitou Nei & 10 & 18 & 93 & 109 & 158 & 278 & 326 & 358 & 340 & 359 & 381 \\
        
        Studier Keppler & 13 & 21 & 99 & 118 & 157 & 211 & 353 & 354 & 346 & 359 & 379 \\
        
        UNJ & 10 & 18 & 94 & 117 & 164 & 229 & 330 & 370 & 359 & 400 & 412 \\
        \hline
    \end{tabular}
    \caption{Peak memory usage in megabytes for ten to one thousand profiles using the lazy version.}
    \label{table:lazymemory}
\end{table}

The difference in the results of the eager and lazy versions can be better compared in individual plot graphs. Figures \ref{graph:mstmemory}, \ref{graph:gcpmemory} and \ref{graph:njmemory} represent these differences for \ac{MST}, \ac{GCP} and \ac{NJ} algorithms respectively.

\begin{figure}[!ht]
    \centering
    \begin{subfigure}[b]{0.49\textwidth}
        %
    \end{subfigure}
    \caption{Peak memory usage in megabytes for MST algorithms compared to their memory complexity.}
    \label{graph:mstmemory}
\end{figure}

\begin{figure}[!ht]
    \centering
    \begin{subfigure}[b]{0.49\textwidth}
        %
    \end{subfigure}
    \caption{Peak memory usage in megabytes for GCP algorithms compared to their memory complexity.}
    \label{graph:gcpmemory}
\end{figure}

\begin{figure}[!ht]
    \centering
    \begin{subfigure}[b]{0.49\textwidth}
        %
    \end{subfigure}
    \caption{Peak memory usage in megabytes for NJ algorithms compared to their memory complexity.}
    \label{graph:njmemory}
\end{figure}

From these individual plot graphs it is possible to see that the implementations of the \ac{MST} and \ac{GCP} algorithms have a memory complexity of $\mathcal{O}(n^2)$, while the implementations of the \ac{NJ} algorithms tend more towards a memory complexity of $\mathcal{O}(n)$. However, despite that, the implementations of the \ac{NJ} algorithms seem to have an overhead great enough to still require more memory than the implementations of the \ac{MST} and \ac{GCP} algorithms with smaller datasets. It is also possible to see that the memory results do not follow a clear pattern and the difference between the lazy and eager versions is almost inexistent. Thus, reusing a distance matrix stored in a file is almost insignificant in terms of memory for all implemented algorithms.

\section{Discussion}

The results obtained from the experimental evaluation lead to the conclusion that the implementations of the algorithms conform to their theoretical time complexity. That is, the implementations of the \ac{NJ} algorithms have a time complexity of $\mathcal{O}(n^3)$, while the implementations of the \ac{MST} and \ac{GCP} algorithms have a time complexity of $\mathcal{O}(n^2)$. These results also show that the Edmonds algorithm is clearly slower than the other \ac{MST} and \ac{GCP} algorithms, despite having the same time complexity.

In terms of memory performance, the obtained results lead to the conclusion that the implementations of the \ac{MST} and \ac{GCP} algorithms have a memory complexity of $\mathcal{O}(n^2)$, while the implementations of the \ac{NJ} algorithms tend more towards a memory complexity of $\mathcal{O}(n)$. Additionally, these results show that the implementations of the \ac{NJ} algorithms seem to have an overhead great enough to still require more memory than the implementations of the \ac{MST} and \ac{GCP} algorithms with smaller datasets.

From these results, it is possible to see a difference in running time between the eager and lazy versions of the algorithms. However, these differences are hardly noticeable, due to the small impact that the distance calculation step has on the workflow, compared to the inference algorithm step, except when using the \ac{goeBURST} algorithm, as its running times are small enough to notice a significant difference. And, despite not following a clear pattern, it is also possible to see that the difference between the lazy and eager versions in terms of memory usage is almost inexistent. Thus, the benefit of storing the distance matrix in a file and reusing it is almost insignificant in terms of running time and memory usage for most algorithms.

\chapter{Final Remarks}

This chapter provides the final remarks for this document, namely a summary of conclusions regarding the whole project, including a summary of the phylogenetic analysis workflow, the objectives of this project, and the time and memory evaluation results. This chapter also provides an enumeration of future work that can be implemented on top of this project to both extend and improve it even further.

\section{Conclusions}

The phylogenetic analysis workflow can be summarized into four consecutive steps, the distance calculation, distance correction, inference algorithm, and local optimization steps. The first step consists of producing a distance matrix from a dataset, including several sequences, through a distance calculation method, such as Hamming, GrapeTree, or Kimura, that calculates the distances between each pair of sequences of the dataset. The dataset can be represented in several formats, including \ac{MLST}, \ac{MLVA}, FASTA, and \ac{SNP}. The second step takes a distance matrix and corrects each distance using a correction formula, such as Jukes-Cantor. This step is optional, thus it may be skipped. The third step transforms a distance matrix into a phylogenetic tree by running a clustering algorithm, such as \ac{goeBURST}, GrapeTree, \ac{UPGMA}, or \ac{NJ} by Studier and Keppler. The phylogenetic tree can be represented in several formats, including Newick and Nexus. And the fourth step takes a phylogenetic tree and tries to locally optimize it through a local optimization algorithm, such as \ac{LBR}. This step is also optional, thus it may be skipped, however it may also be applied several times.

The goal of this project was to develop a command line application that conforms to the phylogenetic analysis workflow and is highly performant, extensible, reusable, and portable. It is different from other existing tools in the sense that it was built to be continuously extended and not just serve a single purpose. It enables reading datasets, distance matrices, and phylogenetic trees from files, calculating and correcting a distance matrix, inferring and locally optimizing a phylogenetic tree, and writing distance matrices and phylogenetic trees to files. Additionally, it provides the capabilities of executing only certain steps of the workflow as well as outputting the results of each step, which can be used as a way to stop and resume the workflow whenever the user desires. This is another thing that other tools do not offer, yet is particularly useful in certain scenarios, such as when the user intends to run several inference algorithms over the same input data, but does not wish to waste time or resources computing the same distance matrix for all of them.

The time performance benchmarks of the experimental evaluation show that the implementations of the algorithms conform to their theoretical time complexity, namely $\mathcal{O}(n^3)$ for \ac{NJ} algorithms and $\mathcal{O}(n^2)$ for \ac{MST} and \ac{GCP} algorithms. However, the implementation of the Edmonds algorithm was shown to have a considerable overhead compared to other \ac{MST} and \ac{GCP} algorithms, despite having the same time complexity. Meanwhile, the memory performance benchmarks lead to the conclusion that the implementations of the \ac{MST} and \ac{GCP} algorithms have a memory complexity of $\mathcal{O}(n^2)$, while the implementations of the \ac{NJ} algorithms tend more towards a memory complexity of $\mathcal{O}(n)$. However, despite showing that the implementations of the \ac{NJ} algorithms have a lower memory complexity, it is also shown that they seem to have an overhead great enough to still require more memory than the implementations of the \ac{MST} and \ac{GCP} algorithms with smaller datasets.

From the results obtained in the experimental evaluation, it is possible to see a difference in running time between the eager and lazy versions of the algorithms. However, these differences are hardly noticeable due to the small impact that the distance calculation step has on the workflow, compared to the inference algorithm step, except when using the \ac{goeBURST} algorithm, as its running times are small enough to notice a significant difference. And, despite not following a clear pattern, it is also possible to see that the difference between the lazy and eager versions in terms of memory usage is almost inexistent. Thus, the benefit of storing the distance matrix in a file and reusing it is almost insignificant in terms of running time and memory usage for most algorithms.

\section{Future Work}

The result of this project boils down to a library that is efficient, reusable, extensible and portable. However, it can still be further extended to include more distance and correction metrics, inference and local optimization algorithms, and dataset, distance matrix and phylogenetic tree formats. Furthermore, it can still be extended in other ways, namely by including other optional steps in the phylogenetic analysis workflow, such as the dynamic addition of relationships between the inference algorithm and local optimization steps, and the calculation of visualization coordinates after all other steps. Also, despite it already being efficient, its time and memory performances can still be improved upon by, for example, introducing parallelization in the algorithms and a cache system in the distance matrix.

\renewcommand{\bibname}{References}
\addcontentsline{toc}{chapter}{\bibname}
\bibliographystyle{unsrturl}
\bibliography{basics/references.bib}

\end{document}